\documentclass{aa}  
\usepackage{natbib}
\bibpunct{(}{)}{;}{a}{}{,}
\usepackage{comment} 
\usepackage{graphicx}
\usepackage{xcolor}
\usepackage{txfonts}
\usepackage{hyperref}
\hypersetup{
    colorlinks,
    linkcolor={blue},
    citecolor={blue},
    urlcolor={blue}
}
\usepackage{tabularx}

\usepackage[noabbrev,capitalize,nameinlink]{cleveref}
\begin{document}

   \title{Episodic outbursts during brown dwarf formation}
   %\subtitle{MRI-driven luminosity outbursts}
   \author{Adam Parkosidis\inst{1},
            Dimitris Stamatellos\inst{2}
          \& 
            Basmah Riaz\inst{3}
          }

   \institute{Anton Pannekoek Institute for Astronomy, University of Amsterdam, Amsterdam 1098 XH, NL\\
              \email{a.parkosidis@uva.nl}
         \and
             Jeremiah Horrocks Institute for Mathematics, Physics \& Astronomy, University of Central Lancashire, Preston PR1 2HE, UK\\         
        \and 
             Universit\"{a}ts-Sternwarte M\"{u}nchen, Ludwig Maximilians Universit\"{a}t, Scheinerstra$\beta$e 1, 81679 M\"{u}nchen, Germany\\
             }

   \date{Received ...; accepted 27/02/2025}

% \abstract{}{}{}{}{} 
% 5 {} token are mandatory
 
  \abstract
   {}
   %It is uncertain whether the formation and evolutionary stages of sub-stellar mass, i.e. brown dwarfs and planetary-mass objects, are comparable to those of stars.
   {There is evidence that stars and browns dwarfs grow through episodic rather than continuous gas accretion. However, the role of episodic accretion in the formation of brown dwarfs remains mostly unexplored. We investigate the role of episodic accretion, triggered by the magnetorotational instability in the inner disk regions, resulting in episodic outbursts during the formation of brown dwarfs, and its implications for their early formation stages.}
   {We use hydrodynamical simulations coupled with a sub-grid  accretion model to investigate the formation of young proto-brown dwarfs and protostars, taking into account the effects of episodic accretion resulting in episodic radiative feedback, i.e. in luminosity outbursts.}
   {The formation timescale for proto brown dwarfs is at least one order of magnitude shorter than that of protostars. Episodic accretion leads to shorter main accretion phase compared to continuous accretion in brown dwarfs, whereas the opposite is true for low-mass stars. Episodic accretion can accelerate early mass accretion in proto-brown dwarfs and protostars, but it results in less massive objects by the end of the main phase compared to continuous accretion. We find an approximately linear correlation between an object’s mass at the end of the main accretion phase and the timing of the last episodic outburst: later events result in more massive brown dwarfs but less massive low-mass stars. Episodic outbursts have a stronger effect on brown dwarf-forming cloud cores, with the last outburst essentially splitting the brown dwarf evolution into a short high-accretion and a much longer low-accretion phase.}
   %explaining why observed proto-brown dwarfs often show lower accretion rates than protostars.
  {}
  %{}

   \keywords{accretion, accretion disks – hydrodynamics – radiative transfer – brown dwarfs - stars: formation
               }

   \maketitle

\section{Introduction}\label{sec:one}

Star formation occurs in dense cores within molecular clouds. Low-mass protostars grow in mass by accreting gas from the collapsing molecular cloud. Since the contracting cloud contains a substantial amount of angular momentum, the infalling gas forms an accretion disk around the protostar \citep{shu1987star}. If a reliable viscosity-generating mechanism exists to transport angular momentum outwards in the disk, the disk's material will spiral inward, while gravitational energy is being transformed to radiation and internal energy. Mechanisms that can provide a viscosity in the disk are the gravitational instability (GI) \citep[e.g.][]{lodato2004testing} and the magneto-rotational instability (MRI) \citep{RevModPhys.70.1}. 

Despite this broad picture, a detailed physical understanding of the accretion process during the proto-stellar stage, when the forming star is still embedded within and accreting from its parent dense core, remains elusive. In the simplest model of star formation, the collapse of a singular isothermal sphere, mass accretes onto the protostar at a constant rate of $\sim 2 \times 10^{-6}$ M$_{\odot}$ yr$^{-1}$ \citep{shu1987star}. However, as \cite{1990AJ.....99..869K} first discussed, most protostars have luminosities that are significantly lower than the accretion luminosity expected from accretion at this rate. This luminosity problem has been highlighted by Spitzer Space Telescope \citep{2004ApJS..154....1W} observations of nearby star-forming regions, which reveal a significant population of protostars with luminosities lower than these theoretical predictions \citep[e.g.][]{2009ApJS..181..321E,2015ApJS..220...11D}.

There is growing evidence that episodic rather than continuous mass accretion is the solution to the luminosity problem, with long periods of very low accretion punctuated by short bursts of rapid accretion (resulting in luminosity outbursts) \citep{2014prpl.conf..387A, Herczeg:2017a,Mercer:2020a, 2023ASPC..534..355F}. Indeed, the luminosities during the periods of low accretion predicted by simulations \citep{2012ApJ...747...52D,stamatellos2012episodic,2015ApJ...805..115V} match the observed proto-stellar luminosity distribution and can alleviate the luminosity problem.

An open question in star formation is whether the formation and evolutionary stages of sub-stellar mass, i.e. brown dwarfs (BDs) (0.013 M$_{\odot}\leq {\rm M} \leq 0.08$ M$_{\odot}$) and planetary-mass objects  ($ \leq 0.013$ M$_{\odot}$) are comparable to those of low-mass stars. In order to unravel this mystery, a combination of theoretical models and observations of young BDs in their most embedded stages of their formation (comparable to the Class 0/I stages of low-mass star formation, here called proto-BDs) is essential. Despite the fact that numerical simulation studies have provided insight to the various brown dwarf formation processes, e.g. whether it is a star-like formation via gravitational core collapse \citep{machida2009first} or alternative mechanisms of formation via disk fragmentation or as ejected embryos \citep{goodwin2007brown,stamatellos2009properties, bate2012stellar,Mercer:2017a,Mercer:2020a},  the number of observed proto-BDs candidates remains small. 

 Observations of Class 0/I brown dwarf candidates \citep{palau2014ic,2015ApJ...807...55M,riaz2015very, 2024MNRAS.529.3601R} suggest that the formation of these systems can be explained by a scaled-down version of star formation. In more recent years, telescopes such as ALMA have allowed us to spatially resolve the inner few au scales of these very low luminosity objects (VeLLOs) and constrain theoretical models \citep{2018ApJ...865..131L}. Observations have revealed features, such as infalling envelopes, rotationally supported pseudo-disk, and outflow-driving area \citep{riaz2017first,riaz2019alma,riaz2021complex}, that are similar to those found in proto-stellar systems. The bolometric luminosity of these objects is generally $\leq 0.1$ L$_{\odot}$, while mean accretion rates range from $\sim 10^{-6}$ to $\sim 10^{-8}$ M$_{\odot}$  yr$^{-1}$ \citep{2021MNRAS.501.3781R}.

Observations of the formation of proto-BDs are quite challenging. The physical scales of early-stage Class 0/I brown dwarfs are estimated to be at least 10 times smaller than those of low-mass protostars \citep{machida2009first,riaz2019alma}. Furthermore, the main accretion phase, i.e. the duration of the Class 0 and Class I phases combined, differs significantly between brown dwarfs and low-mass stars. The envelope dissipation timescale is predicted to be quite brief, $\lesssim 10^4$ yr, in the formation of brown dwarfs, according to simulations performed by \cite{machida2009first,2012MNRAS.421..588M}. In contrast, protostars with ages of less than $0.1$ Myr have been shown observable signs of infall from an envelope \citep{2012ApJ...754...52T,2011ApJ...732...97D}, while \cite{2015ApJS..220...11D} estimate  lifetimes  of $0.13$-$0.26$ Myr  for the Class 0 phase and $0.27$-$0.52$ Myr for the Class I phase.

In most previous theoretical  studies, a continuous gas accretion is assumed onto the proto-BD. However, in the inner few au the gas can be hot enough to be thermally ionized and couple with magnetic fields activating the MRI \citep[e.g.][]{2009ApJ...694.1045Z,2010ApJ...713.1134Z}. As a result, the accretion can be episodic switching between periods of enhanced and quiescent accretion when the MRI is active or not, respectively, just as it happens on solar-mass stars. Episodic accretion may also be due to infalling fragments formed due to GI \cite[e.g.][]{2005ApJ...633L.137V,2010ApJ...719.1896V,2015ApJ...805..115V}, or due to thermal instability in the inner disk region \citep{Bell:1995a}.

\cite{Stamatellos:2011a, stamatellos2012episodic} examine the role of episodic accretion on disk fragmentation around young solar-mass stars, regulating the number of low-mass stars and BDs forming \citep[see also][]{Mercer:2017a}. \cite{Lomax:2014a,Lomax:2015a} show that episodic accretion is necessary for reproducing the initial mass function (IMF) of stellar clusters, as continuous accretion leads to a lack of low-mass stars and BDs. \cite{MacFarlane:2017a} \& \cite{Vorobyov:2020l} study the effect of episodic outbursts on the structures of protostellar disks. 
\cite{MacFarlane:2019a, MacFarlane:2019p} discuss the expected observational signatures of episodic accretion/outbursts around young stars.

The role of episodic accretion during the formation of proto-BD remains mostly unexplored. Observations of multiple shocked emission knots at the extended jet (HH1165) of the very low-luminosity Class I candidate proto-BD Mayrit 1701117 \citep{riaz2017first, riaz2019alma} indicate that episodic accretion events may also happen during brown dwarf formation.

In this paper, we use the Smoothed Particle Hydrodynamics (SPH) code {\sc Seren} \citep{hubber2011seren} to simulate the effect of episodic accretion/episodic outbursts during brown dwarf formation via the gravitational collapse of a very low-mass pre-BD core. We probe the evolution of the proto-BD until the end of the main accretion phase \citep{machida2009first}, examining the role of the timescale of outbursts and the interval between them. We discuss the observational implications of our models and compare our results with the proto-brown dwarf observations of \cite{riaz2021complex}. Additionally, we compare the results of simulations of BD formation with those of low-mass star formation.

\section{Computational method}\label{sec:two}

We use the Smoothed Particle Hydrodynamics code SEREN \citep{hubber2011seren} to treat the gas hydro- and thermo- dynamics. The code uses a 2nd-order Runge-Kutta integration approach, multiple particle time-steps, and an octal tree to compute gravitational forces and find particle neighbors. In order to decrease artificial shear viscosity, the code employs time-dependent artificial viscosity \citep{morris1997switch} with parameters $\alpha_{\rm min}= 0.1$, $\beta= 2\alpha$, and a Balsara switch \citep{balsara1995neumann}. 

The \cite{stamatellos2007radiative} technique is used to treat the chemical and radiative processes that control gas temperature \citep[see also][]{forgan2009introducing,Mercer:2018}.  The technique accounts for radiative cooling/heating, viscous heating, background radiation field heating, and compressional heating. The equation of state takes into account the rotational and vibrational degrees of freedom of molecular hydrogen, as well as the various chemical states of hydrogen and helium  \citep{boley2007internal}. It is assumed that the gas is a mixture of $70\%$ hydrogen and $30\%$ helium. The dust and gas opacity is set to $\kappa(\rho, T) = \kappa_{0}\rho^{\alpha} T^{\beta}$, where $\kappa_{0},\alpha, \beta$ are constants \citep{Bell:1995a} that depend on the physical processes causing the opacity, such as dust sublimation, ice mantle melting, molecular contributions, and $H^{-}$ contributions.

The self-gravitating gas dynamics, energy equation, and related radiation transport are treated explicitly up to densities of $\rho = 10^{-9}$ g cm$^{-3}$. It is assumed that a gravitationally confined condensation with $\rho > 10^{-9}$ g cm$^{-3}$ will eventually develop into a bound object (protostar or proto-BD). At this point, a sink \citep{bate1995modelling}, with a radius of $1$ au, replaces the condensation to avoid very short time-steps. Only the sink's gravity and luminosity affects the rest of the computational domain. Gas, that flows within the sink radius and is bound to it, is added to the sink's mass.  A forming proto-BD will provide radiative feedback to the surrounding environment, setting the pseudo-background radiation field, i.e. a minimum temperature floor under which the gas cannot cool radiatively \citep[for details see][]{stamatellos2007radiative, Mercer:2017a} that is set to 
\begin{equation}\label{eq:pseudo-background_radiation}
    T^4_{\rm BGR}(\vec{r}) = (10K)^4 + \frac{L}{16\pi\sigma_{SB}|\vec{r} - \vec{r_{\rm BD}}|^2} \; ,
\end{equation}
where $L$ and $\vec{r}_{\rm sink}$ are the luminosity and the position of the proto-BD, respectively.  The luminosity at the early stages that we model is due to gas accretion onto the object, and is given by:
\begin{equation}\label{eq:accretion_luminosity}
    L=  f \frac{GM_{\rm BD}\dot{M}_{BD}}{R_{BD}} \; ,
\end{equation}
where $M_{\rm BD}$ is the mass of the proto-BD, $R_{\rm BD}$ its radius and $\dot{M}_{\rm BD}$ the accretion rate onto it. $f=0.75$ is the fraction of the accretion energy that is radiated away at the photosphere of the proto-BD, rather than being expended in e.g. driving jets and/or winds \citep{2007prpl.conf..261S,2010ApJ...725.1485O,stamatellos2012episodic}. We assume $R_{\rm BD} = 3$ R$_{\odot}$ which is the typical radius of a young forming object \citep{1993ApJ...418..414P,2011MNRAS.417.2036B}.  The background temperature profile is variable (see Eqs. \ref{eq:pseudo-background_radiation} and  \ref{eq:accretion_luminosity}). It depends on the accretion rate onto the proto-BD and on the distance from it, $T_{\rm BGR}  \propto r^{-2}$, which is the anticipated radial temperature profile on an envelope around a luminosity source.

\section{Model setup}\label{sec:three}

\subsection{A semi-analytical model of episodic accretion}

We use the model of \cite{Stamatellos:2011a} to include the effect of episodic accretion (resulting in episodic radiative feedback, i.e. in luminosity outbursts) on a forming proto-BD. The model uses a sub-grid model to describe the region within a sink SPH particle. Here, we briefly describe the basic aspects of the model that are needed to interpret the results of our simulations. More details of the model are given in \cite{Stamatellos:2011a,stamatellos2012episodic}.  We assume that the MRI develops in the inner region ($<1$~AU) around BDs the same way it does around solar-mass stars. It is therefore assumed that the density and temperature conditions very close to the BDs are appropriate for this to happen \citep{2014ApJ...795...61B}.

The model assumes that the sink consists of the proto-BD ($M_{\rm BD}$) and an inner accretion disk (IAD; $M_{\rm IAD}$) which is notionally the part of the disk that is not resolved by the SPH simulation. Gas that is accreted onto the sink, first is deposited on the IAD and then spirals towards the proto-BD according to the local disk viscosity.  When the viscosity is low, gas piles up in the IAD until the temperature rises to the point where thermal ionization couples the matter to the magnetic field. The temperature at which this occurs is set to $T_{\rm MRI} \sim 1400 K$. At this point, the MRI is triggered, moves angular momentum outward and the IAD gas spirals inward and onto the central proto-BD. The accretion rate onto the proto-BD is 
\begin{equation}\label{eq:star_accretion_rate}
    \dot{M}_{\rm BD} = \dot{M}_{\rm BGR} + \dot{M}_{\rm MRI} \; ,
\end{equation}
where $\dot{M}_{\rm BGR}$ is the quiescent accretion rate from the IAD to the proto-BD, whereas $\dot{M}_{\rm MRI}$ is a significantly greater accretion rate that only occurs when the MRI operates. 

The MRI is activated when enough mass has accumulated in the IAD, i.e.
\begin{equation}\label{eq:IAD_mass_condition}
    M_{\rm IAD} > M_{\rm MRI} \sim \dot{M}_{\rm MRI} \Delta t_{\rm MRI} \; ,
\end{equation}
where
\begin{equation}\label{eq:MRI_accretion_rate}
    \dot{M}_{\rm MRI} \sim 5 \times 10^{-4} \text{M$_{\odot}$ yr$^{-1}$} \Bigl(\frac{\alpha_{\rm MRI}}{0.1}\Bigr) \;.
\end{equation}
$\alpha_{\rm MRI}$ is the effective Shakura-Sunyayev parameter \citep{1973A&A....24..337S} for the MRI-provided viscosity, and $\Delta t_{\rm MRI}$
is the duration of the episodic event
\begin{equation}\label{eq:MRI_event_duration}
    \begin{split}
    \Delta t_{\rm MRI} & \sim 0.25 \text{kyr}\ \Bigl(\frac{0.1}{\alpha_{\rm MRI}}\Bigr) \Bigl(\frac{M_{\rm BD}}{0.2 \text{M$_{\odot}$}}\Bigr)^{2/3}  \\
    & \times \Bigl(\frac{\dot{M}_{\rm IAD}}{10^{-5} \text{M$_{\odot}$ yr$^{-1}$}}\Bigr)^{1/9} \; ,
    \end{split}
\end{equation}
where $\dot{M}_{\rm IAD}$ is the accretion rate onto the sink and onto the IAD. 

Using Equations \eqref{eq:IAD_mass_condition}, \eqref{eq:MRI_accretion_rate} and \eqref{eq:MRI_event_duration}  the condition for the MRI activation becomes
\begin{equation}\label{eq:IAD_mass_condition_2}
    M_{\rm IAD} > 0.13 \text{M$_{\odot}$} \Bigl(\frac{M_{\rm BD}}{0.2 \text{M$_{\odot}$}}\Bigr)^{2/3} \Bigl(\frac{\dot{M}_{\rm IAD}}{10^{-5} \text{M$_{\odot}$ yr$^{-1}$}}\Bigr)^{1/9} \; .
\end{equation}
To match observations of FU Ori-type outbursts the accretion rate during the outburst is set to exponentially drop as 
\begin{equation}\label{eq:MRI_accretion_rate_2}
\begin{split}
    \dot{M}_{\rm MRI} &=  7.9  \times 10^{-4} \text{M$_{\odot}$ yr$^{-1}$} \Bigl(\frac{\alpha_{\rm MRI}}{0.1}\Bigr)\\
    & \times \exp\Bigl(-\frac{t - t_0}{\Delta t_{\rm MRI}}\Bigr), \; t_0 < t < \Delta t_{\rm MRI}\,,
    \end{split}
\end{equation}
where $t_0$ marks the start of the event. 
After the IAD empties, the time taken to re-accumulate enough gas for another episodic accretion event is  $\Delta t_{acc} \sim  M_{\rm MRI}/\dot{M}_{IAD}$, thus
\begin{equation}\label{eq:time_interval_outburst}
    \Delta t_{acc} = 13 \text{kyr}\ \Bigl(\frac{M_{\rm BD}}{0.2 \text{M$_{\odot}$}}\Bigr)^{2/3} \Bigl(\frac{\dot{M}_{\rm IAD}}{10^{-5} \text{M$_{\odot}$ yr$^{-1}$}}\Bigr)^{-8/9} .
\end{equation}

There are two free parameters in the model: (i) the viscosity parameter provided by the MRI, $\alpha_{\rm MRI}$, and (ii) the quiescent accretion rate, $\dot{M}_{\rm BGR}$. The first, regulates the intensity and duration of the outburst; raising $\alpha_{\rm MRI}$ results in a shorter (Eq.~\ref{eq:MRI_event_duration}), more intense outburst (Eq.~\ref{eq:accretion_luminosity}). $\dot{M}_{\rm BGR}$ takes over as the dominant mode of gas accretion onto the proto-BD when the MRI is inactive (Eq.~\ref{eq:star_accretion_rate}). The values of $\alpha_{\rm MRI}$ and $\dot{M}_{\rm BGR}$ are uncertain.

\subsection{BD formation - Initial conditions}

We simulate the collapse of a low-mass  molecular cloud core so that the central object forming is a BD rather than a star. The initial density profile of the cloud is
\begin{equation}\label{eq:dens_prof}
    \rho(r) = \frac{\rho_{\rm c}}{1+(r/R_{\rm kernel})^2}\,,
\end{equation}
where $\rho_{\rm c} = 2 \times 10^{-14}$  g cm$^{-3}$ is the central density and $R_{\rm kernel} = 21$ au is the radius of the core  where the density is roughly uniform. The core extends to $R_{\rm core} = 500$ au and its total mass is $M_{\rm core} = 0.087$ M$_{\odot}$.  The initial ratio of the rotational to gravitational energy is $\beta_{\rm rot} = U_{\rm rot}/|U_{\rm grav}|=0.01$. The initial temperature of the gas is $T_{\rm gas} = 10$ K and thus the initial ratio of thermal to gravitational energy is $\alpha_{\rm thermal} = U_{\rm thermal}/|U_{\rm grav}|=0.41$.  We assume that the cloud core does not have any turbulence.

\begingroup
%\onecolumngrid
\begin{figure*}[!htbp]
    \centering
    \includegraphics[width=\linewidth]{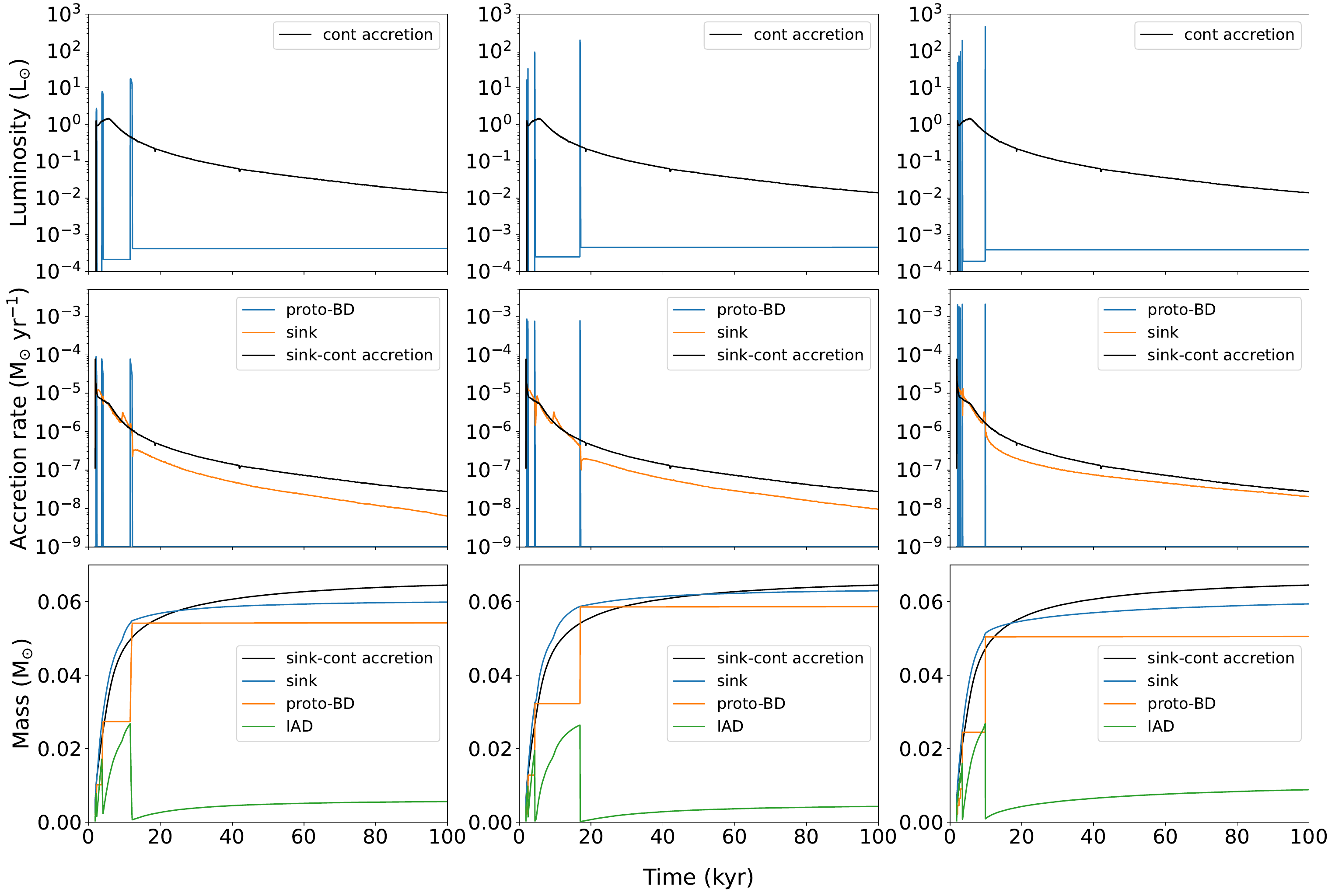}
    \caption{ Luminosity of the central object, accretion rate onto the sink and proto-BD, and the masses of the proto-BD, IAD and sink (i.e. proto-BD + IAD mass) (top to bottom row) for runs 1-3 (left to right column);  see also Table~\ref{tab:runs}. In black, the luminosity, the accretion rate onto the sink and its mass for run 7. } %The quiescent accretion rate is $\dot{M}_{\rm BGR} = 10^{-9}$ M$_{\odot}$  yr$^{-1}$.}
    \label{fig:a_mri_model}
\end{figure*}

\begin{figure*}[!htbp]
    \centering
    \includegraphics[width=\linewidth]{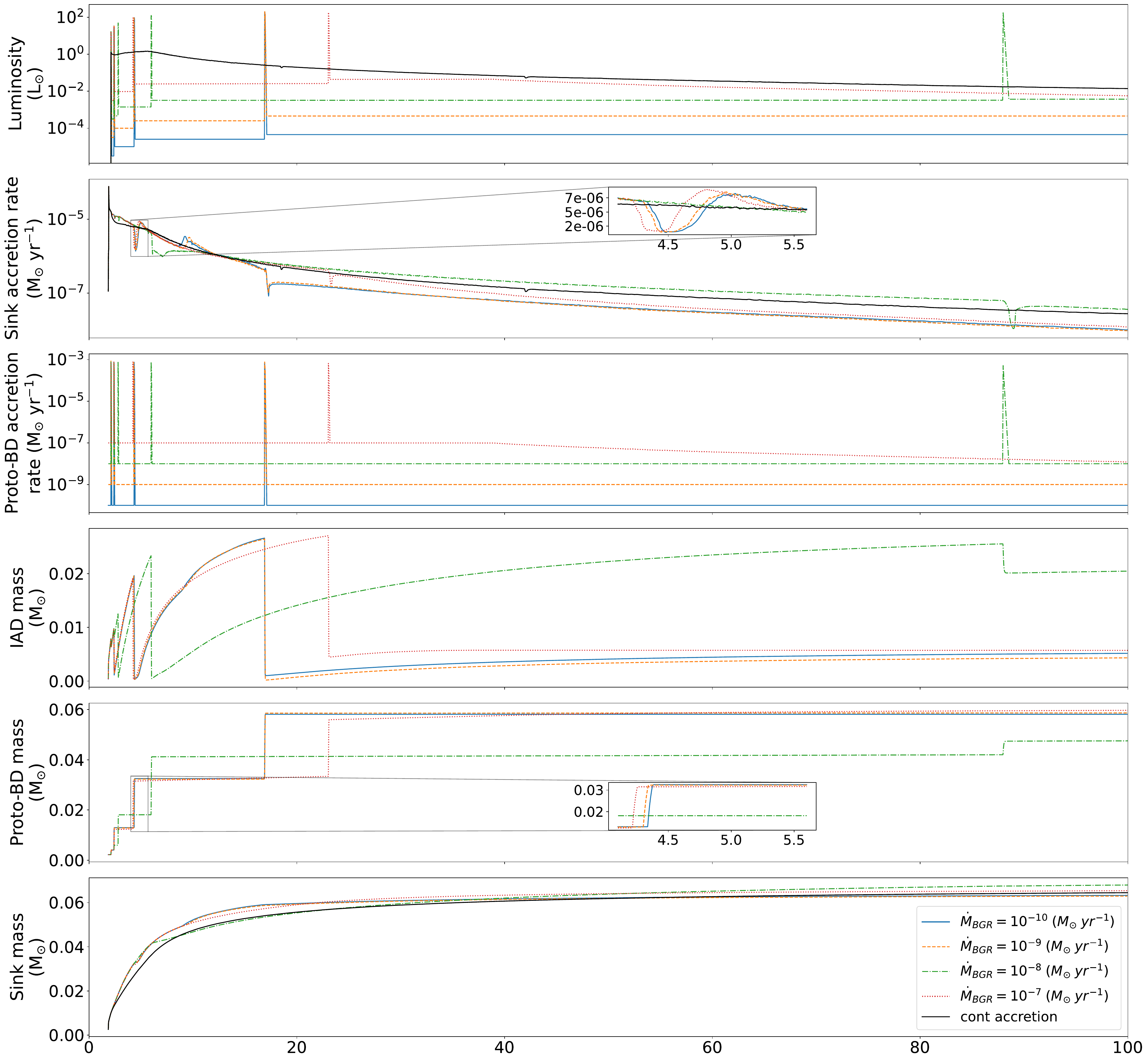}
    \caption{ Luminosity of the central object, accretion rate onto the sink and proto-BD, and the masses of the proto-BD, IAD and sink (i.e. proto-BD + IAD mass) for runs 2,4-6 (top to bottom row);  see also Table~\ref{tab:runs}. In black, the luminosity, the accretion rate onto the sink and its mass for run 7.}
    \label{fig:bgr_model}
\end{figure*}
\endgroup
%\twocolumngrid

The cloud core is represented by $3 \times 10^5$ SPH particles, thus each particle has mass $m_{\rm particle} = 3 \times 10^{-7}$ M$_{\odot}$. The minimum mass that can be resolved is therefore $M_{\rm min} = N_{\rm neib}m_{\rm particle} = 1.8 \times 10^{-5}$ M$_{\odot}$. This is two orders of magnitude smaller than the opacity limit for fragmentation, i.e. the theoretical minimum mass for star formation, $\sim 10^{-3}$ M$_{\odot}$ \citep[e.g.][]{2006A&A...458..817W}, thus the self-gravitating condensations are well resolved.

We perform seven simulations with the same cloud initial conditions.  In six of them, we vary the parameters $a_{\rm MRI}$ and $\dot{M}_{\rm BGR}$, which control the episodic accretion/episodic outburst events. In the remaining one, the episodic accretion model is inactive, i.e. the accretion and the associated radiative feedback are assumed to be continuous. Observations and simulations of protoplanetary disks around low-mass stars suggest $\alpha_{\rm MRI} \in [0.01,0.4]$ \citep{2007MNRAS.376.1740K,2009ApJ...701..260I,2014ApJ...795...61B,2020ApJ...895...41K,2021MNRAS.504..280J}. In this work, we choose $\alpha_{\rm MRI} = [0.01,0.1,0.3]$.  The accretion rate onto Class0/I proto-BD candidates \citep{palau2014ic,riaz2015very,2018ApJ...865..131L,2021MNRAS.501.3781R} are thought to be $\dot{M}_{\rm BGR} \in [10^{-6}, 10^{-9}]$ M$_{\odot}$~yr$^{-1}$, thus we chose four values in this range. In Table~\ref{tab:runs_parameters}, we summarize the initial conditions of our simulations.

\begin{table}[!htbp]
\caption{Simulation parameters for BD formation.}
  \centering
  \begin{tabular}{cccccc}
  \hline
  run & $\dot{M}_{\rm BGR}$  & $\alpha_{\rm MRI}$ & $\alpha_{\rm thermal}$  & $\beta_{\rm rot}$\\
  & (M$_{\odot}$ yr$^{-1}$) & & & \\
  \hline \hline
  1 & $10^{-9}$ & 0.01 & 0.41 & 0.01\\
  2 & $10^{-9}$ & 0.1 & 0.41 & 0.01\\
  3 & $10^{-9}$ & 0.3 & 0.41 & 0.01\\
  4 & $10^{-10}$ & 0.1 & 0.41 & 0.01\\
  5 & $10^{-8}$ & 0.1 & 0.41 & 0.01\\
  6 & $10^{-7}$ & 0.1 & 0.41 & 0.01 \\
  7 & – & – & 0.41 & 0.01 \\
  \hline
  \end{tabular}\label{tab:runs}
\label{tab:runs_parameters}
\tablefoot{ $\dot{M}_{\rm BGR}$ is the quiescent accretion rate onto the forming proto-BD, and $\alpha_{\rm MRI}$ is the viscosity $\alpha$-parameter for the MRI. $\alpha_{\rm thermal}$ is the initial thermal to gravitational energy and $\beta_{\rm rot}$ is the initial rotational to gravitational energy.
}
\end{table}

\section{Episodic accretion during brown dwarf formation}\label{sec:four}

All simulations run for $t = 100$ kyr. In Fig.~\ref{fig:a_mri_model}, we present the results of the simulations for different $\alpha_{\rm MRI}$ (runs 1-3), and in Fig.~\ref{fig:bgr_model} the results of the simulations for different $\dot{M}_{\rm BGR}$ (runs  2, 4-6). In both figures, we also include the results of the simulation with continuous accretion, i.e. where the episodic accretion model is inactive (run 7). Note that, the proto-BD and the IAD are not defined for run 7. 

The collapsing cloud core forms a sink at $t \sim 1.85$ kyr, hence a proto-BD with an inner accretion disk is acquired. The sink's initial mass is $M_{\rm sink,1.85kyr} =0.0026$ M$_{\odot}$. At first, the sink accretes vigorously, with $\dot{M}_{\rm IAD} \sim 7.8 \times 10^{-5}$ M$_{\odot}$  yr$^{-1}$, but the accretion rate decreases as the available cloud mass is reduced. Following the formation of a sink, gas begins to accumulate in the IAD before accreting onto the proto-BD. When MRI is inactive, the forming proto-BD accumulates mass at a constant $\dot{M}_{\rm BGR}$ rate. Once the MRI is triggered, $\dot{M}_{\rm BD}$ rises up by $\sim 4-6$ orders of magnitude, and the IAD deposes intensively its mass on to the forming proto-BD (see Figs.~\ref{fig:a_mri_model} and \ref{fig:bgr_model}). After the end of the event, the IAD fills up until the next episodic accretion event.

Accretion on to the proto-BD releases energy (Eq. \ref{eq:accretion_luminosity}). When the MRI is inactive, the forming object's luminosity remains constant and $\leq 0.1$ L$_{\odot}$. However, during episodic accretion events, the luminosity dramatically increases and then declines exponentially until the event ends. The young proto-BD exhibits outbursts during the first few kyr of its life, but no outbursts occur after $t = 20$ kyr, because the majority of the cloud core mass has already been accreted onto the forming object. The exception is run 5, where the last outburst occurs at $t \sim 80$ kyr. 
 It is therefore expected that outbursts from BDs happen within a relatively short timescale, making their discovery difficult.

\begin{figure}[!htbp]
    \centering
    \includegraphics[width=\linewidth]{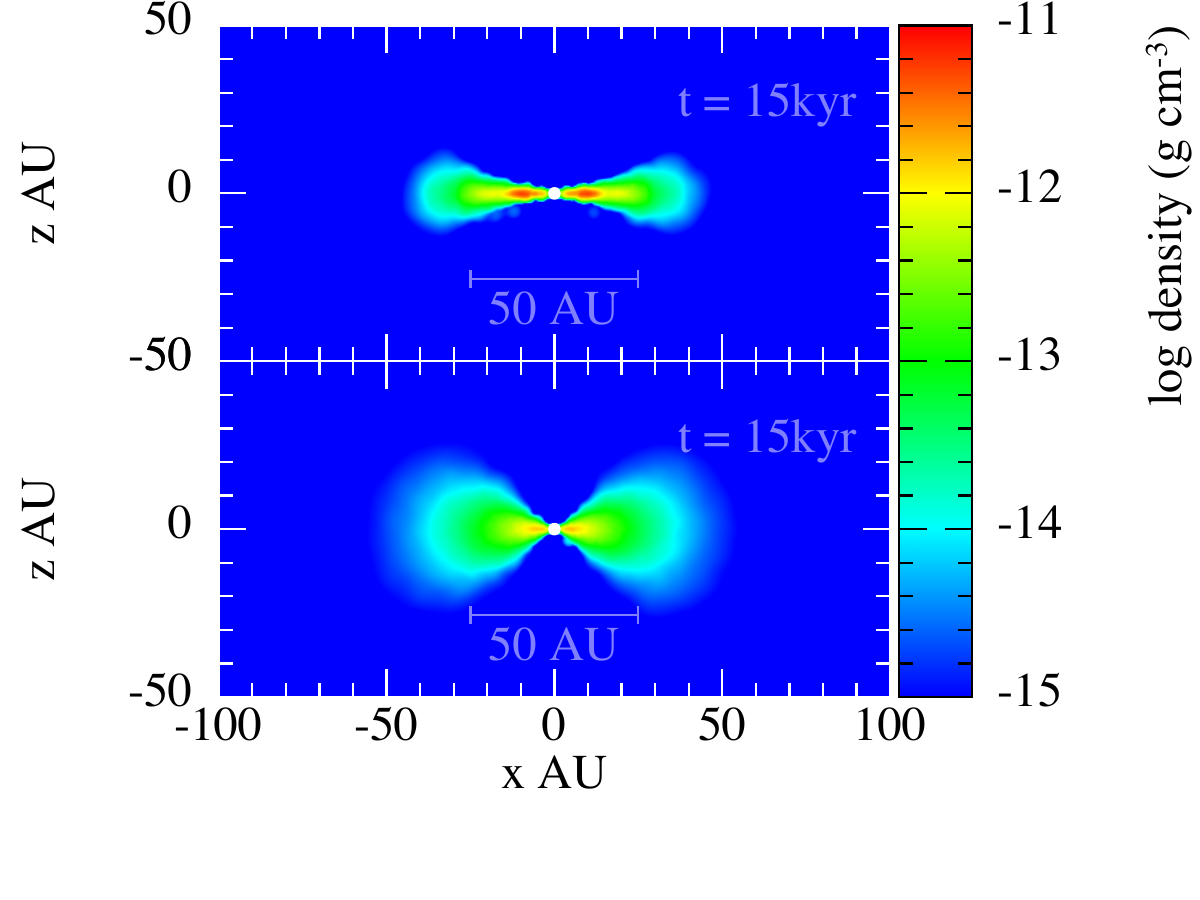}
    \caption{Cross-section of the density on the $x-z$ plane showing the disc  (edge-on view) at $t=15$ kyr, which forms around the proto-BD in run 2 (episodic accretion; top) and in run 7 (continuous accretion; bottom).}
    \label{fig:bd-discs}
\end{figure}

In the continuous accretion case (run 7) the evolution of gas accretion is different. Initially, the sink accretes at similar rates as in other models, and the accretion rate gradually declines as the available cloud mass decreases. However, in runs with episodic accretion (1-6) the accretion rate onto the sink shows a significant drop after the last outburst, and during the subsequent evolution (except for run 5). Furthermore, in the continuous accretion run the object displays a higher luminosity than models 1-6, a pattern that is temporarily disrupted when MRI is active and an outburst occurs. Fig.~\ref{fig:bd-discs} illustrates two snapshots of the discs around the forming BD for run 2 and run 7, respectively. For continuous accretion, the higher luminosity results in a hotter and thus thicker disc. Nevertheless, the total energy emitted during each run is on the same order of magnitude, $E_{\rm total}\sim 10^{45}$ erg.  Episodic accretion distributes the total emitted energy into short bursts, allowing for long periods of low luminosities of proto-brown dwarfs. 

Radiative feedback directly affects the accretion onto the sink, because it heats the gas,  leading to  a thicker disc. Consequently, the surface density drops (Fig.~\ref{fig:bd-discs}) and accretion is delayed or even temporary suppressed \citep[e.g.][]{Mercer:2017a}. This is evident in Fig.~\ref{fig:bgr_model} (2$^{\rm nd}$ row, inset), where the accretion rate onto the IAD/sink drops during each outburst, but after the radiative feedback is terminated, the gas cools down quickly and the accretion rate increases again. For a longer episodic feedback, the accretion onto the sink decreases for a longer period of time. As a result, the duration and the time between successive outbursts affects the sink's growth during the early stage of formation. In the next subsections, we explore the effect of the models' parameters on the duration and the time intervals between successive outbursts, and the overall evolution of the BD.

\subsection{The effect of increasing $a_{\rm MRI}$}

\begin{figure}[!htbp]
    \centering
    \includegraphics[width=\linewidth]{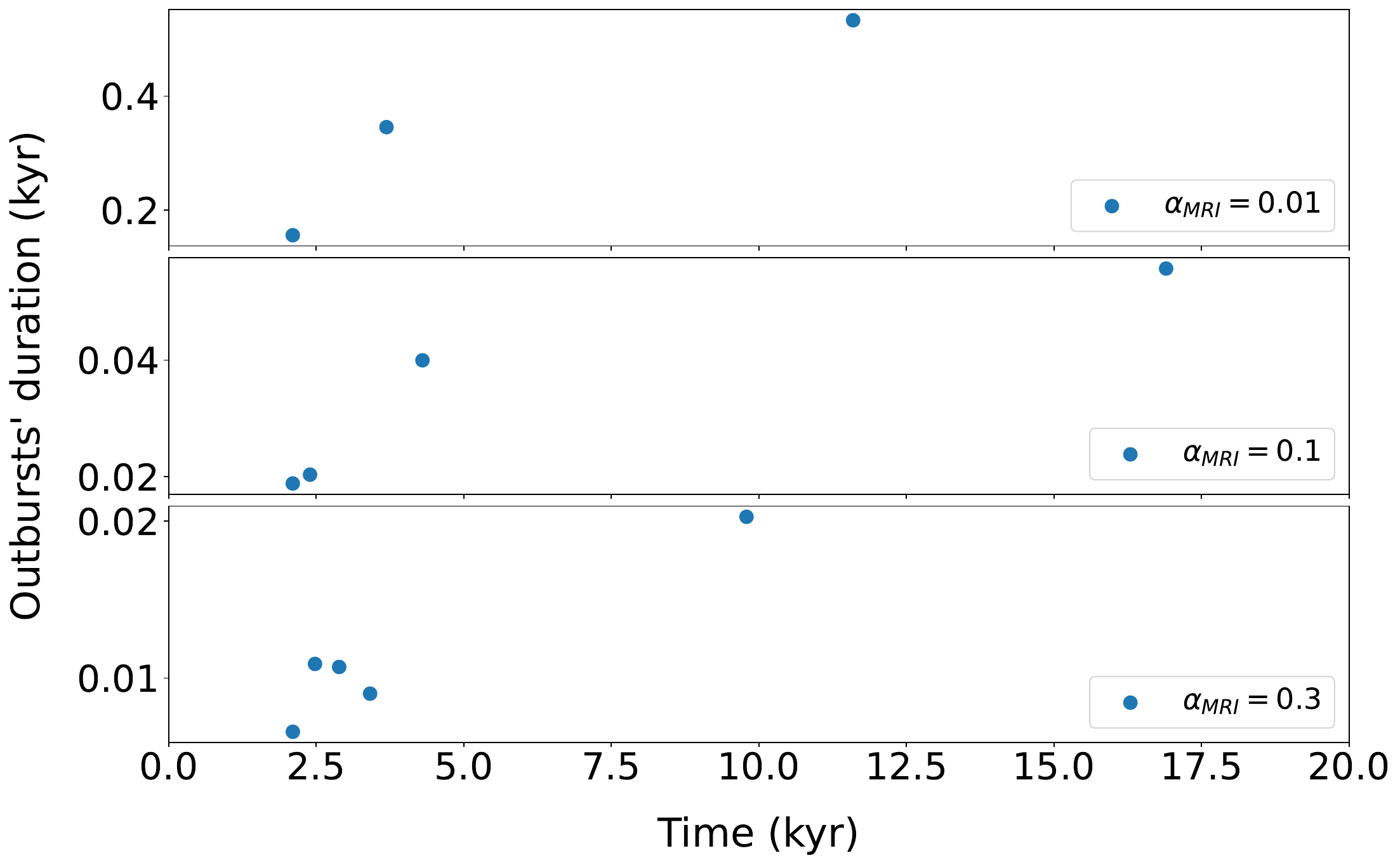}
    \caption{Effect of increasing $\alpha_{\rm MRI}$ on the duration of episodic outbursts for runs 1-3,  see also Table~\ref{tab:runs}.}
    \label{fig:duration_same_bgr}
\end{figure}

In Figure~\ref{fig:duration_same_bgr}, we present the duration of successive outbursts for different viscosity parameter $\alpha_{\rm MRI}$ values during the first $20$ kyr of the simulation, as no episodic event occurs after $\sim 17$ kyr. The duration of the episodic events decreases, with increasing $\alpha_{\rm MRI}$. As briefly explained in Sec.~\ref{sec:three}, when the MRI is active, the viscosity parameter $\alpha_{\rm MRI}$ regulates how quickly matter accretes onto the proto-BD; the greater $\alpha_{\rm MRI}$, the more quickly angular momentum can be transferred outward through the IAD and the faster matter accretes onto the proto-BD. For $\alpha_{\rm  MRI} = 0.01$ the duration of the shortest is $ \sim 156$ yr; while, for $\alpha_{\rm MRI} = 0.3$, the duration of the lengthiest event is just $ \sim 20$ yr. Consequently, the duration of the respective outbursts decreases as $\alpha_{\rm MRI}$ increases (see also Eq. \ref{eq:MRI_event_duration}).

\begin{figure}[!htbp]
    \centering
    \includegraphics[width=\linewidth]{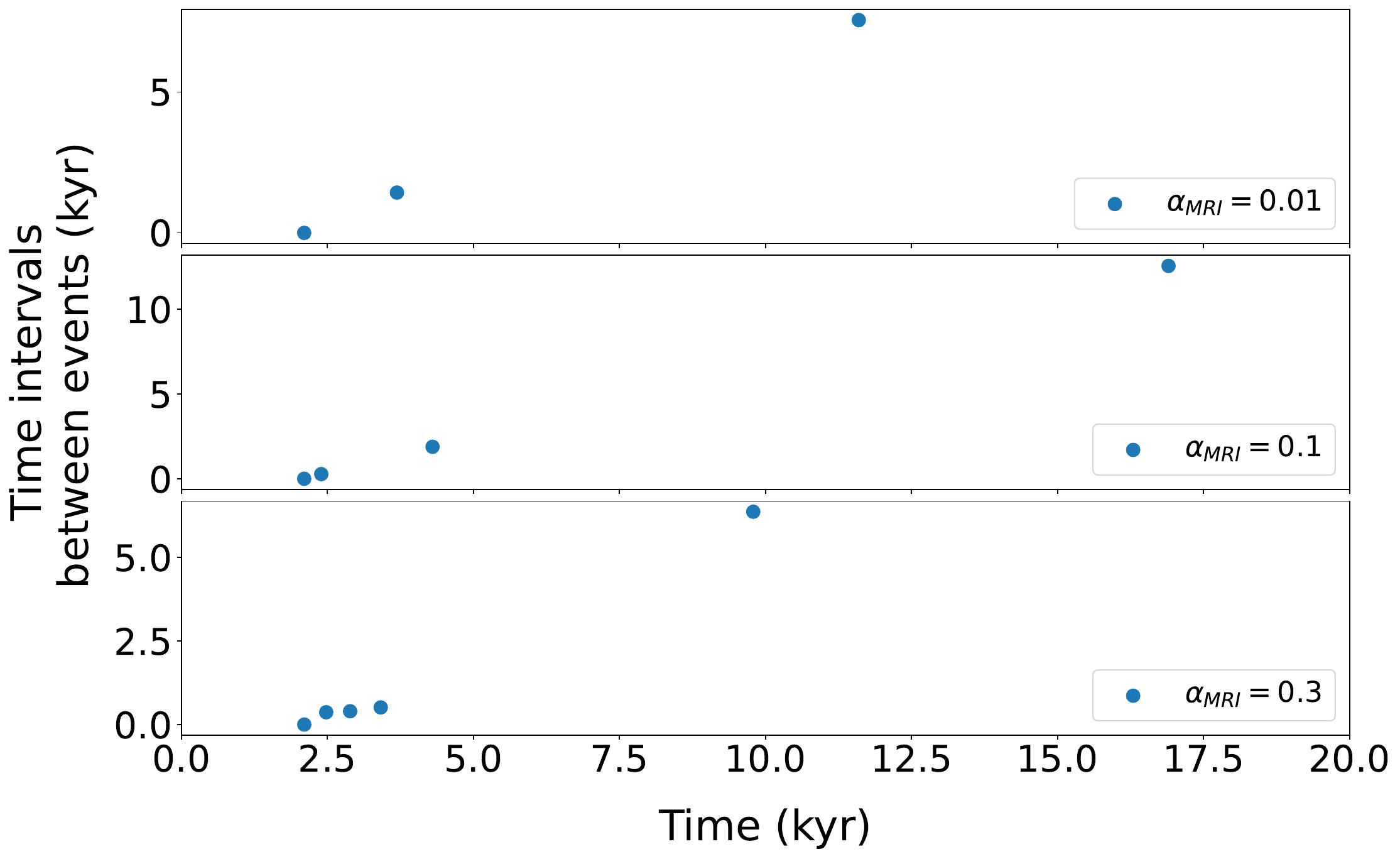}
    \caption{ Effect of increasing $\alpha_{\rm MRI}$ on the time intervals between successive episodic outbursts for runs 1-3; see also Table~\ref{tab:runs}.}
    \label{fig:time_intervals_same_bgr}
\end{figure}

In Figure~\ref{fig:time_intervals_same_bgr}, we plot the time intervals between successive outbursts. For a given $\alpha_{\rm MRI}$ value, these intervals increase as the system evolves. This is because the mass of the proto-BD (M$_{\rm BD}$) increases over time, while the accretion rate onto the sink ($\dot{M}_{\rm IAD}$) decreases (Eq.~\ref{eq:time_interval_outburst}). 

Additionally, with increasing $\alpha_{\rm MRI}$, we observe two trends: (1) the total number of episodic events increases, and (2) the time intervals between successive outbursts decrease. Higher $\alpha_{\rm MRI}$ values are associated with shorter outbursts and, therefore, reduced periods of radiative feedback. This reduction allows the accretion rate onto the IAD/sink ($\dot{M}_{\rm IAD}$) to decrease initially but recover over shorter timescales. As a result, the IAD can re-accumulate gas more rapidly, leading to subsequent bursts on shorter timescales. For instance, the time intervals between the first and the second events are $\sim 1432$ yr and $ \sim 369$ yr, for $\alpha_{\rm MRI} = 0.01$ and $\alpha_{\rm MRI} = 0.3$, respectively. Therefore, shorter outbursts lead to shorter time intervals between successive outbursts, ultimately resulting in a greater number of events.


During an episodic accretion event, the accretion rate onto the proto-BD increases. Higher values of $\alpha_{\rm MRI}$ result to a more significant increase of the accretion rate onto the proto-BD (see 2$^{nd}$ row in Fig.~\ref{fig:a_mri_model} and Eq. \ref{eq:MRI_accretion_rate_2}). For $\alpha_{\rm  MRI} = 0.01$ the accretion rate onto the proto-BD reaches a value of $ \sim 9 \times 10^{-5}$ M$_{\odot} \; yr^{-1}$, while for $\alpha_{\rm MRI} = 0.3$, maximizes at $ \sim 2 \times 10^{-2}$ M$_{\odot} \; yr^{-1}$. Despite the fact, that $\alpha_{\rm MRI}$ determines how quickly matter accretes onto the proto-BD, the amount of mass that is delivered onto the proto-BD is not directly dependent on $\alpha_{\rm MRI}$ (Eq. \ref{eq:IAD_mass_condition_2}); nevertheless, a shorter event duration results to a more intense outburst. Hence, for greater, $\alpha_{\rm MRI}$ the outburst brightness is also higher (Eq. \ref{eq:accretion_luminosity}). For $\alpha_{\rm  MRI} = 0.01$ the luminosity of the proto-BD maximizes at $L_{\rm BD} \sim 18$ L$_{\odot}$ and lasts $\sim 533$ yr (see Fig.~\ref{fig:a_mri_model}). For $\alpha_{\rm MRI} = 0.3$ the luminosity reaches up to $L_{\rm BD} \sim 462$ L$_{\odot}$, but the duration of the event is $<20$ yr, and therefore statistically difficult to be observed.

\subsection{The effect of increasing $\dot{M}_{\rm BGR}$}

The quiescent accretion rate, $\dot{M}_{\rm BGR}$, regulates the growth of the sink and the amount of energy released, when MRI is inactive (see Fig.~\ref{fig:bgr_model} and Eq.~\ref{eq:star_accretion_rate}); higher $\dot{M}_{\rm BGR}$ values, correspond to higher brightness released by the proto-BD. In Figure~\ref{fig:bgr_model}, we see that the sink accretion rates, the IAD and the proto-BD mass converge independently of the $\dot{M}_{\rm BGR}$ value for runs 2,4 and 6, except for run 5.

\begin{figure}[!htbp]
    \centering
    \includegraphics[width=\linewidth]{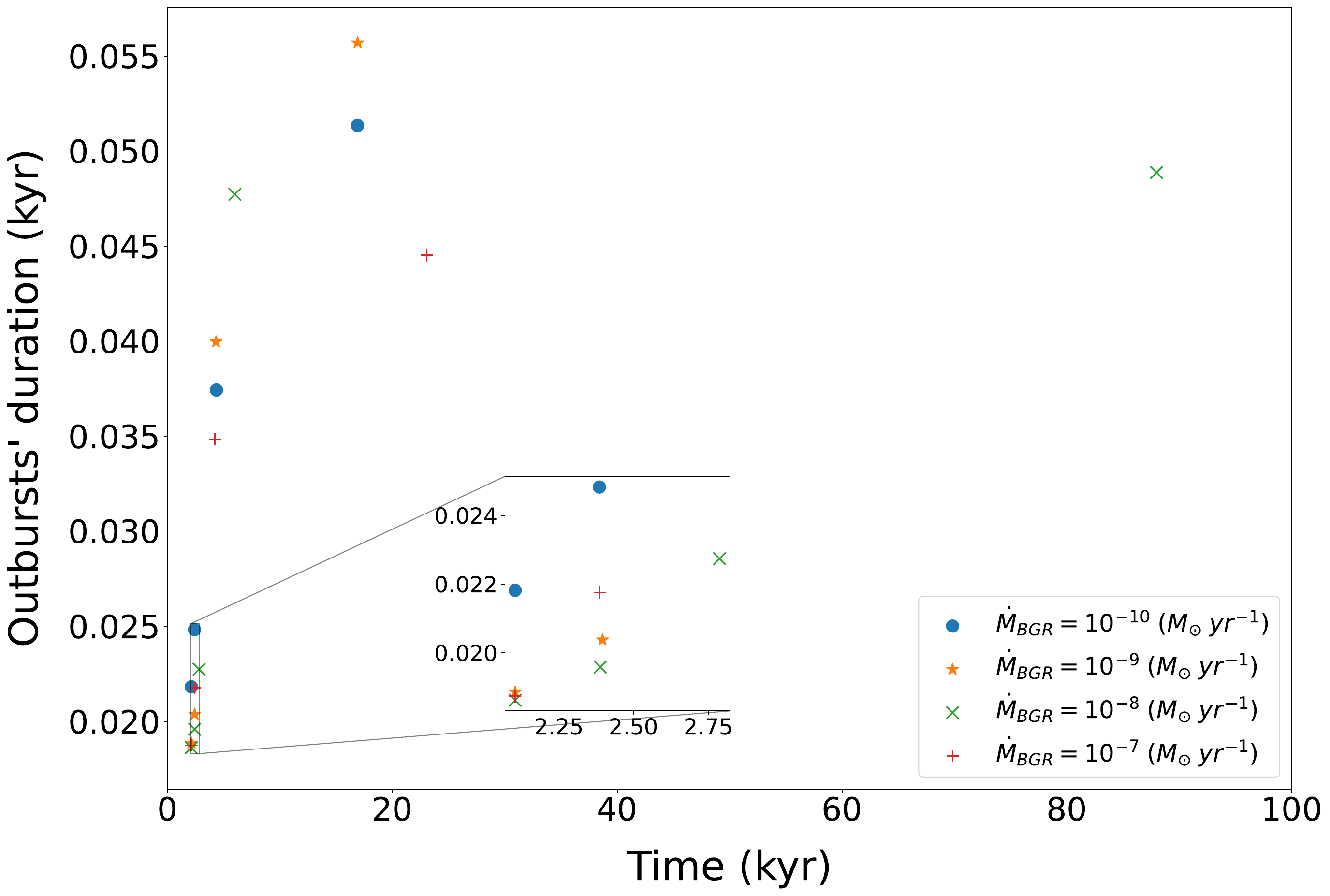}
    \caption{Effect of increasing $\dot{M}_{\rm BGR}$ on the duration of episodic outbursts for runs 2, 4-6;  see also Table~\ref{tab:runs}.}
    \label{fig:duration_same_a}
\end{figure}
\begin{figure}[!htbp]
    \centering
    \includegraphics[width=\linewidth]{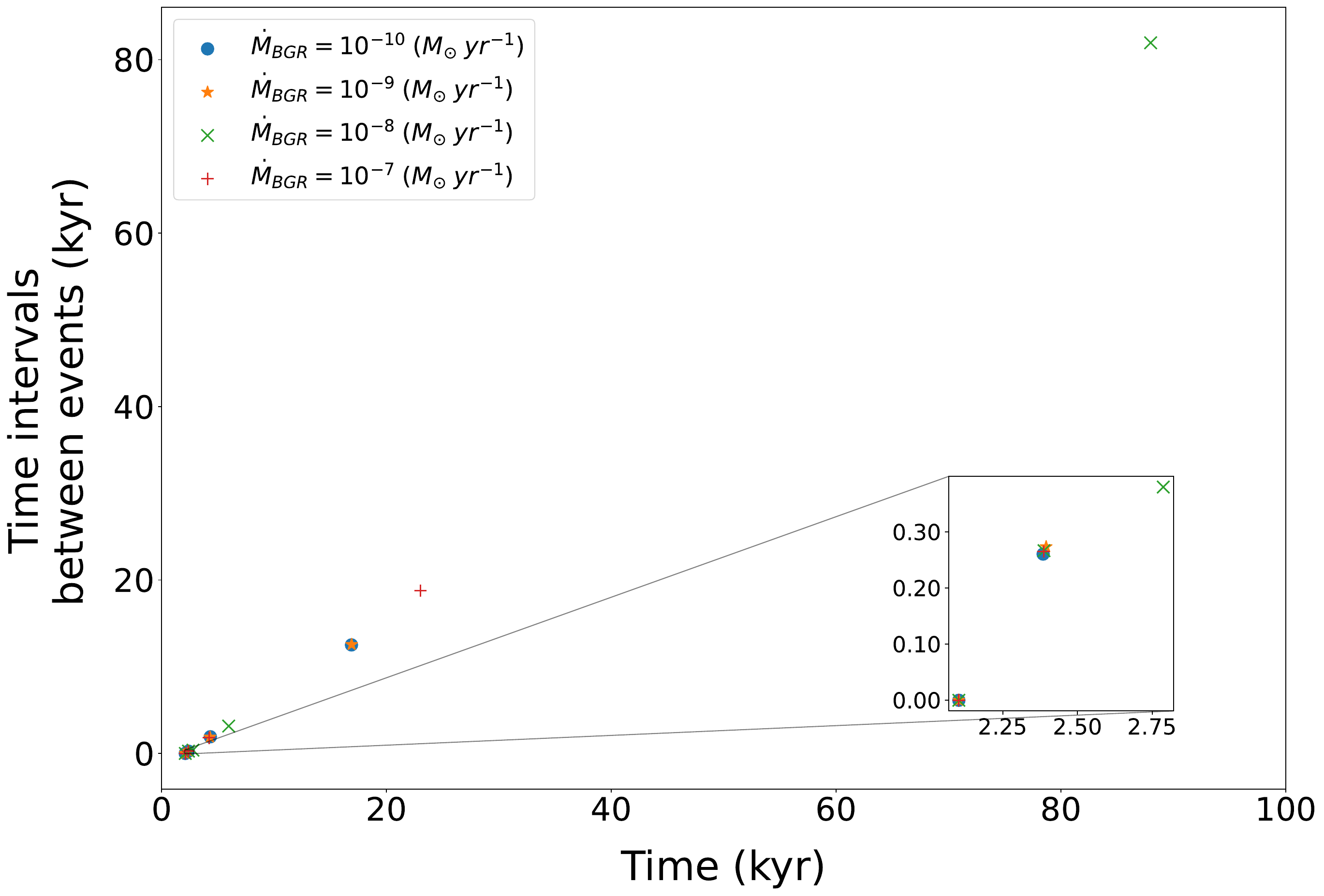}
    \caption{ Effect of increasing $\dot{M}_{\rm BGR}$ on the time intervals between successive episodic outbursts for runs 2, 4-6; see also Table~\ref{tab:runs}.}
    \label{fig:time_intervals_same_a}
\end{figure}

In Figures~\ref{fig:duration_same_a} and \ref{fig:time_intervals_same_a}, we present the duration of successive outbursts and the time intervals between consecutive outbursts for runs 2,4,5 and 6, respectively. The duration of the respective outbursts varies by $\leq 15$ yr for varied $\dot{M}_{\rm BGR}$, and so the duration of the outbursts is mainly independent of $\dot{M}_{\rm BGR}$. The time intervals between successive outbursts vary also by $\leq 15$ yr for varied $\dot{M}_{\rm BGR}$, except for run 5. For the latter, the third outburst, takes place at $\sim 2.8$ kyr, and the subsequent evolution differs from the rest runs (see subset in Fig.~\ref{fig:time_intervals_same_a}). In runs 2,4 and 6, four episodic accretion events occur no later than $\sim 20$ kyr, while in run 5 we encounter five events with the last one taking place at $t \sim 80$ kyr. 

Overall, the duration of episodic outbursts is largely independent of $\dot{M}_{\rm BGR}$, while this parameter may affect the time intervals between episodic outbursts. However, this seems to be indirect and at least one order of magnitude less important than the effect of $\alpha_{\rm MRI}$. In the next sections, we constrain ourselves to the models for different $\alpha_{\rm MRI}$ values, runs 1-3, and we explore the implications of episodic accretion in the formation of BDs.

\subsection{Episodic accretion during the main accretion phase of BD formation}

\begin{figure}[!htbp]
    \centering
    \includegraphics[width=\linewidth]{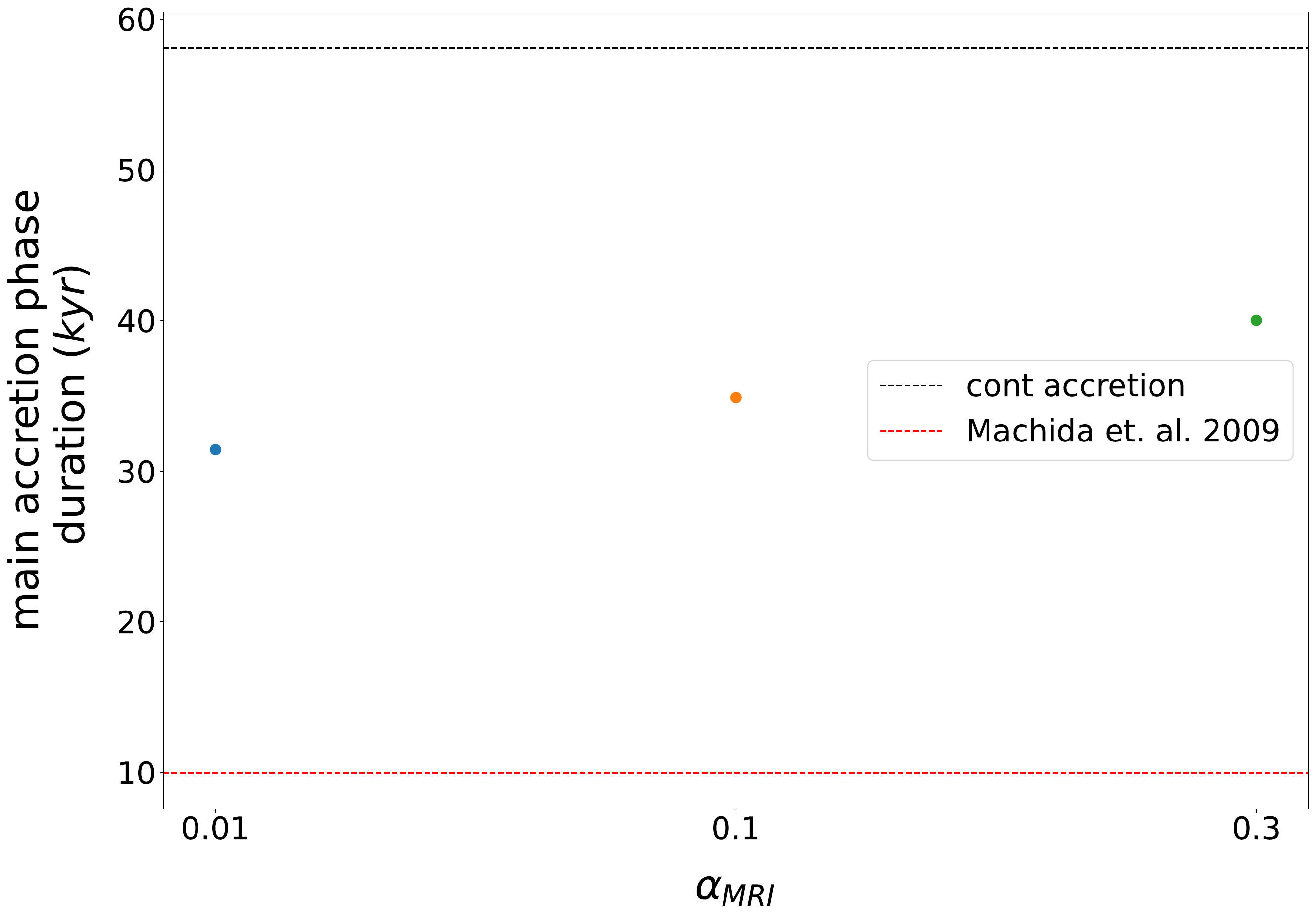}
    \caption{Duration of the main accretion phase for runs 1-3,7; see also Table~\ref{tab:runs}. The black horizontal dashed line corresponds to run 7. The red horizontal dashed line corresponds to the results of \cite{machida2009first}.}
    \label{fig:duration_main_acc_BDs}
\end{figure}

We investigate the implications of episodic accretion in the very early stages of BD formation. We focus on the timescale of the main accretion phase, which corresponds to the Class 0/I stages of low-mass star formation. We define the main accretion phase as the timescale on which the accretion rate onto the sink has decreased by $\sim 3$ orders of magnitude, similar to \cite{machida2009first}. In Figure~\ref{fig:duration_main_acc_BDs}, we present the results for runs 1-3,7. 

The duration of the main accretion phase increases, with increasing $\alpha_{\rm MRI}$, as seen in Fig.~\ref{fig:duration_main_acc_BDs}. Higher $\alpha_{\rm MRI}$ values correspond to shorter but more intense events (see also Fig.~\ref{fig:duration_same_a} and \ref{fig:a_mri_model}). Furthermore, the time intervals between successive events shorten as $\alpha_{\rm MRI}$ increases, while the total number of events increases (see Fig.~\ref{fig:time_intervals_same_a}). As a result, we argue that, despite the larger number of high-accretion events, the presence of less significant but longer ones, combined with prolonged but less intense radiative feedback, leads to shorter timescales for the main accretion phase in BD formation.

Additionally, Fig.~\ref{fig:duration_main_acc_BDs} shows that, regardless of the $\alpha_{\rm MRI}$ values, the main accretion phase of the episodic accretion models 1-3 is shorter than in the continuous accretion model (run 7). In Figure~\ref{fig:a_mri_model}, we see that the sink particles in runs 1-3 gain mass more rapidly (at the expense of the envelope) than the sink in run 7 during the early stages of formation, $\Delta t \leq 20$ kyr. During these early episodic events, $M_{\rm sink} << M_{\rm core}$, and after the radiative feedback from an event ends, the accretion rate onto the sink rises again to values similar to those in the continuous accretion model. However, following the final event, a clear divergence occurs. By this point, the majority of the available mass has been accreted onto the sink, while the remaining low-mass core is heated by radiative feedback causing a significant drop in the accretion rate. Notably, the accretion rate in model 3 rises slowly afterwards, suggesting that there is still a non-negligible amount of gas in the core to be accreted onto the proto-BD. We conclude that episodic accretion during BD formation leads to shorter timescale for the main accretion phase compared to continuous accretion. 

\begin{table}[!htbp]
\caption{Summary of episodic accretion effects on BD formation.}
  \centering
  \begin{tabularx}{\linewidth}{c c c c c c }
  \hline
  run & $t_{\rm main}$ & $t_{\rm final,event}$ & $M_{\rm sink,main }$ & $M_{\rm sink,100 kyr}$ & $\dot{M}_{\rm IAD,main}$\\
  &  (kyr) & (kyr) & (M$_{\odot}$) & (M$_{\odot}$) & (M$_{\odot}$ yr$^{-1}$)\\
  \hline \hline
  1  & 31.43 & 11.6 & 0.0560 & 0.0599 & 7.8 $\times 10^{-8}$ \\
  2  &  34.90 & 16.90 & 0.0592 & 0.0630 & 7.8 $\times 10^{-8}$\\
  3 & 40.01 & 9.80 & 0.0552 & 0.0594 & 7.5 $\times 10^{-8}$\\
  7 & 58.07 & – & 0.0605 & 0.0645 & 7.8 $\times 10^{-8}$\\
  \hline
  \end{tabularx}\label{tab:BD_results}
\tablefoot{$t_{\rm main}$ is the duration of the main accretion phase, while $t_{\rm final,event}$ denotes the moment of the final outburst. $M_{\rm sink,main }$ is the sink mass at the end of the main accretion phase, and $M_{\rm sink,100 kyr}$ is the sink mass at the end of the simulation.  $\dot{M}_{\rm IAD,main}$ indicates the accretion rate into the sink at the end of the main accretion phase.
}
\end{table}

\begin{figure}[!htbp]
    \centering
    \includegraphics[width=\linewidth]{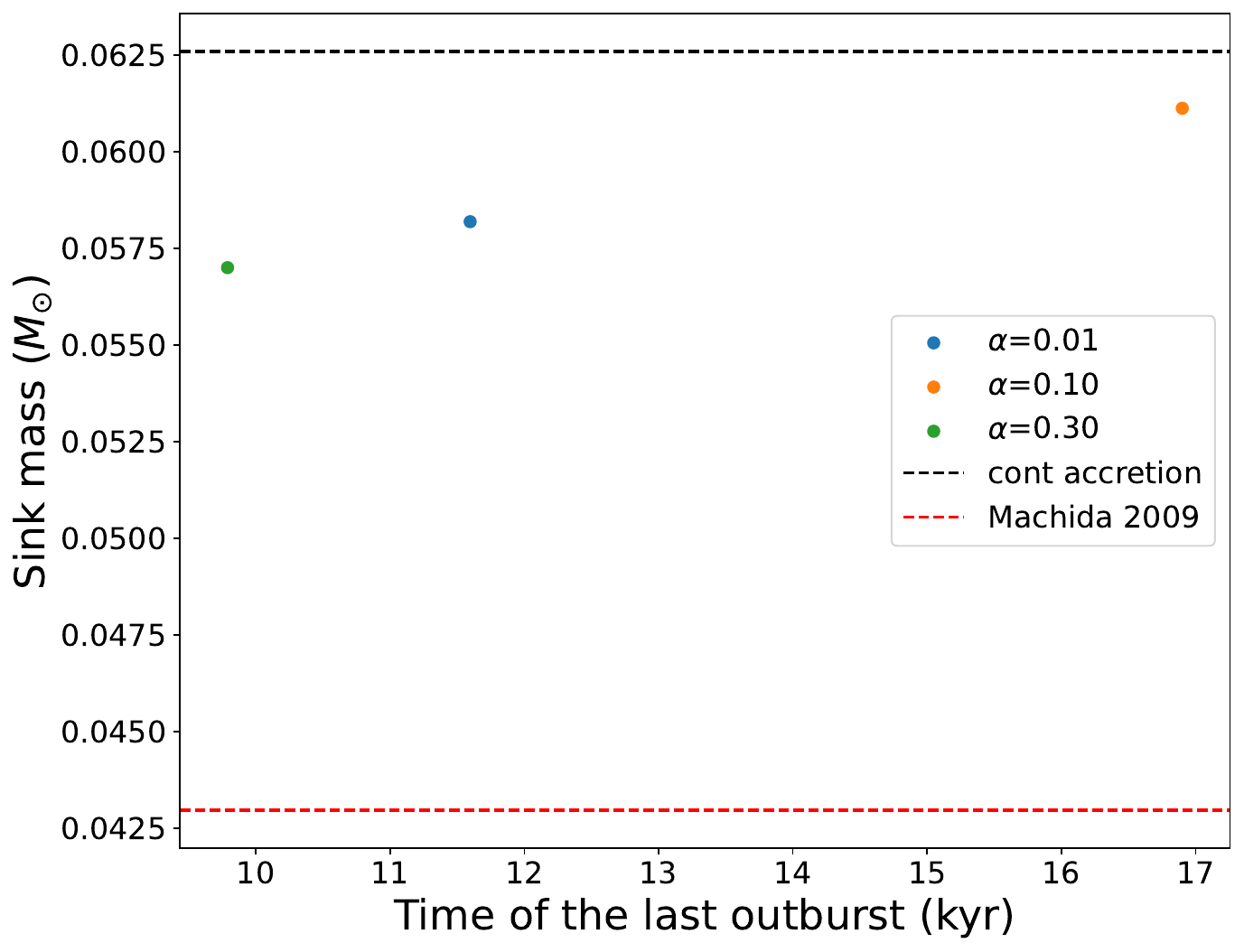}
    \caption{Sink mass at the end of the main accretion phase vs. time of the last outburst for runs 1-3,7;  see also Table~\ref{tab:runs}. The black horizontal dashed line corresponds to run 7 (continuous accretion). The red horizontal dashed line corresponds to the results of \cite{machida2009first}.}
    \label{fig:accreted_mass_main_acc_BDs}
\end{figure}

In Figure~\ref{fig:accreted_mass_main_acc_BDs}, we present sink masses at the end of the main accretion phases as a function of the time of the last episodic event, for runs 1-3. We additionally show the sink mass at the end of the main accretion phase for run 7 and the results of \cite{machida2009first}. The sink particles accumulate more mass when the final episodic event occurs later. Interestingly, this result is also consistent with runs 4–6. Finally, in the continuous accretion model (run 7) the sink accretes more mass than in the episodic accretion models (runs 1–3). In Table~\ref{tab:BD_results} we summarize our results. We conclude that episodic accretion accompanied by radiative outbursts results to lower overall gas accretion onto the proto-BD than continuous accretion.

From Figure~\ref{fig:accreted_mass_main_acc_BDs}, we see an approximately linear correlation between the sink mass at the end of the main accretion phase and the time of the last outburst. The later the final outburst occurs, the more massive the sink at the end of the main accretion phase.

We note that \cite{machida2009first} find a shorter duration for the main accretion phase ($\sim10$~kyr) but they investigate a system that is different to ours, including different physics: a rotating magnetized compact low-mass cloud ($M_{\rm core} = 0.22{\rm M}_{\odot}$). Therefore, such differences are to be expected; their result is used for reference only.

\begingroup
%\onecolumngrid
\begin{figure*}[!htbp]
    \centering
    \includegraphics[width=\linewidth]{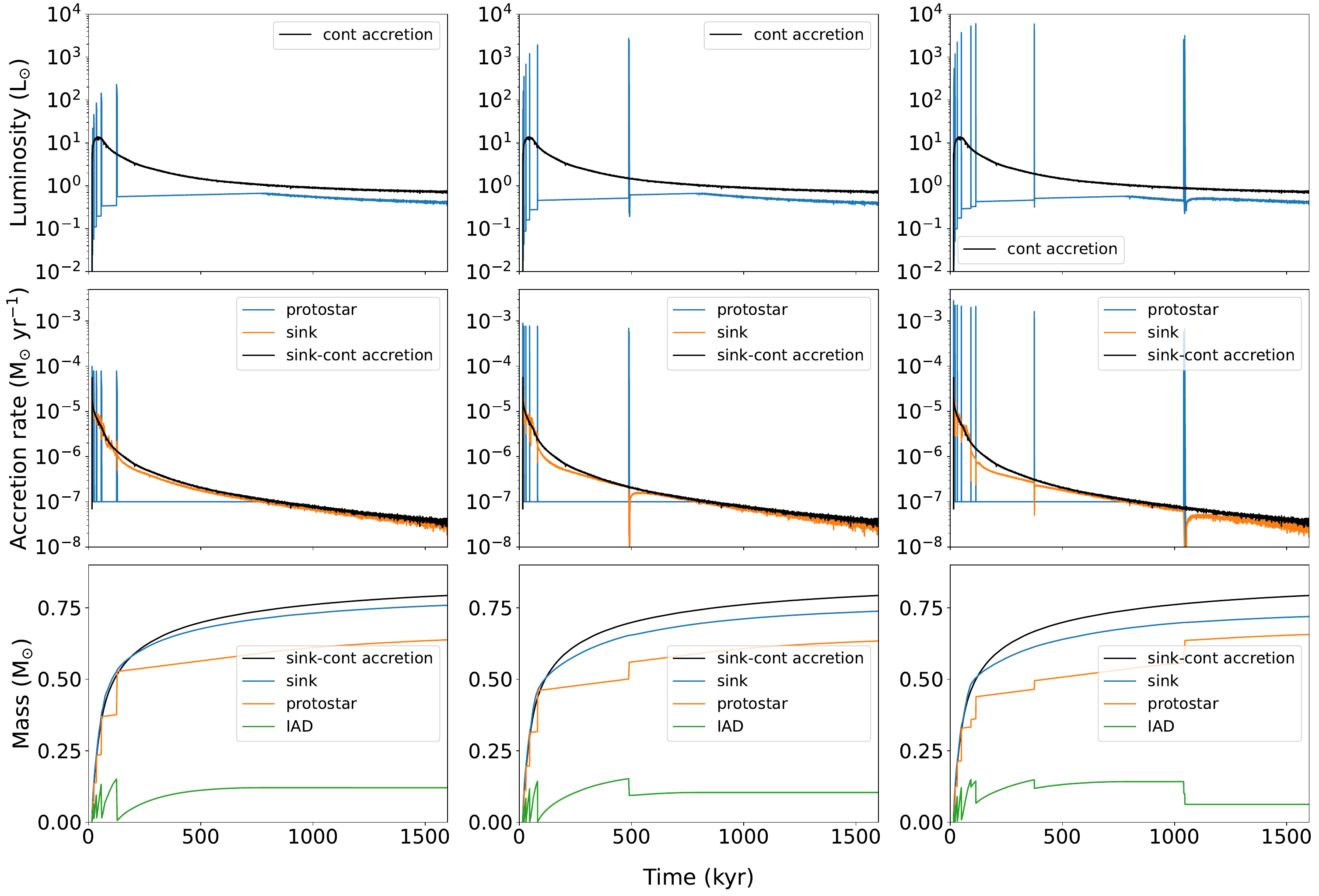}
    \caption{ Luminosity of the central object, accretion rate onto the sink and protostar, and the masses of the protostar, IAD and sink (i.e. protostar + IAD mass) (top to bottom row) for runs 8-10 (left to right column); see also Table~\ref{tab:runs_protostars}. In black, the luminosity, the accretion rate onto the sink and its mass for run 11.}
    \label{fig:a_mri_protostars}
\end{figure*}
\endgroup

%\newpage
%\twocolumngrid

%\section{discussion}\label{sec:five}

\section{Comparison of outbursting proto-BDs with outbursting low-mass protostars}\label{sec:five}

\subsection{Low-mass star formation - Initial conditions}

We investigate the impact of episodic accretion during the early stages of low-mass star formation. We simulate the collapse of a more massive  molecular cloud core so that the forming central object is a protostar, rather than a proto-BD. The cloud core's density profile is given by Eq. \eqref{eq:dens_prof}, we use $\rho_{\rm c} = 2 \times 10^{-16}$ g cm$^{-3}$, $R_{\rm kernel} = 215$ au, and $R_{\rm core} = 5500$ au. Thus, the cloud's core  total mass is $M_{\rm core} = 1.0$~M$_{\odot}$. The initial ratio of the rotational to gravitational energy is $\beta_{\rm rot} = U_{\rm rot}/|U_{\rm grav}|=0.01$. The initial temperature of the gas is $T_{\rm gas} = 10$ K and thus the initial ratio of thermal to gravitational energy is $\alpha_{\rm thermal} = U_{\rm thermal}/|U_{\rm grav}|=0.39$.

The cloud core is represented by $3 \times 10^5$ SPH particles, with each particle having mass $m_{\rm particle} = 3.35 \times 10^{-6}$ M$_{\odot}$. The minimum resolved mass now is $M_{\rm min} = N_{\rm neib}m_{\rm particle} = 2.012 \times 10^{-4}$ M$_{\odot}$, and the self-gravitating condensations are well resolved \citep[e.g.][]{2006A&A...458..817W}.

We preform four simulations of the collapse of this more massive cloud. For runs 8-10, we choose $\dot{M}_{\rm BGR} = 10^{-7}$ M$_{\odot}$ yr$^{-1}$ and $a_{\rm MRI} = 0.01, 0.1, 0.3$, respectively. Run 11 corresponds to a model where  accretion is continuous. In Table~\ref{tab:runs_protostars}, we summarize the initial conditions for the simulations related to low-mass star formation.

\begin{table}[!htbp]
\caption{Simulation parameters for low-mass star formation.}
  \centering
  \begin{tabular}{cccccc}
  \hline
  run & $\dot{M}_{\rm BGR}$  & $\alpha_{\rm MRI}$ & $\alpha_{\rm thermal}$  & $\beta_{\rm rot}$\\
  & (M$_{\odot}$ yr$^{-1}$) & & & \\
  \hline \hline
  8 & $10^{-7}$ & 0.01 & 0.39 & 0.01 \\
  9 & $10^{-7}$ & 0.1 & 0.39 & 0.01 \\
  10 & $10^{-7}$ & 0.3 & 0.39 & 0.01 \\
  11 & – & – & 0.39 & 0.01 \\
  \hline
  \end{tabular}
\label{tab:runs_protostars}
\tablefoot{$\dot{M}_{\rm BGR}$ is the quiescent accretion rate onto the forming protostar, and $\alpha_{\rm MRI}$ is the viscosity $\alpha$-parameter for the MRI. $\alpha_{\rm thermal}$ is the initial thermal to gravitational energy and $\beta_{\rm rot}$ is the initial rotational to gravitational energy.
}
\end{table}

\subsection{Outbursting low-mass protostars}

All simulations run for $t = 1.6$ Myr. In Figure~\ref{fig:a_mri_protostars}, we present the results. We observe qualitatively similar behaviour for protostars as seen for proto-BDs (see Sec.~\ref{sec:four}), though the timescales differ. The collapsing core forms a sink at $t\sim15.26$ kyr, after which a protostar with an inner accretion disk is acquired. The sink's initial mass is $M_{\rm sink,15.26kyr} =0.0048$ M$_{\odot}$. The young protostars undergo 7, 10, and 15 episodic accretion events, accompanied by outbursts, for $a_{\rm MRI} = 0.01, 0.1, 0.3$, respectively. The majority of these events occur within the first $150$ kyr after the sink formation, with the final two and three events happening later for $a_{\rm MRI} = 0.1$ and $0.3$, respectively.

\begin{figure}[!htbp]
    \centering
    \includegraphics[width=\linewidth]{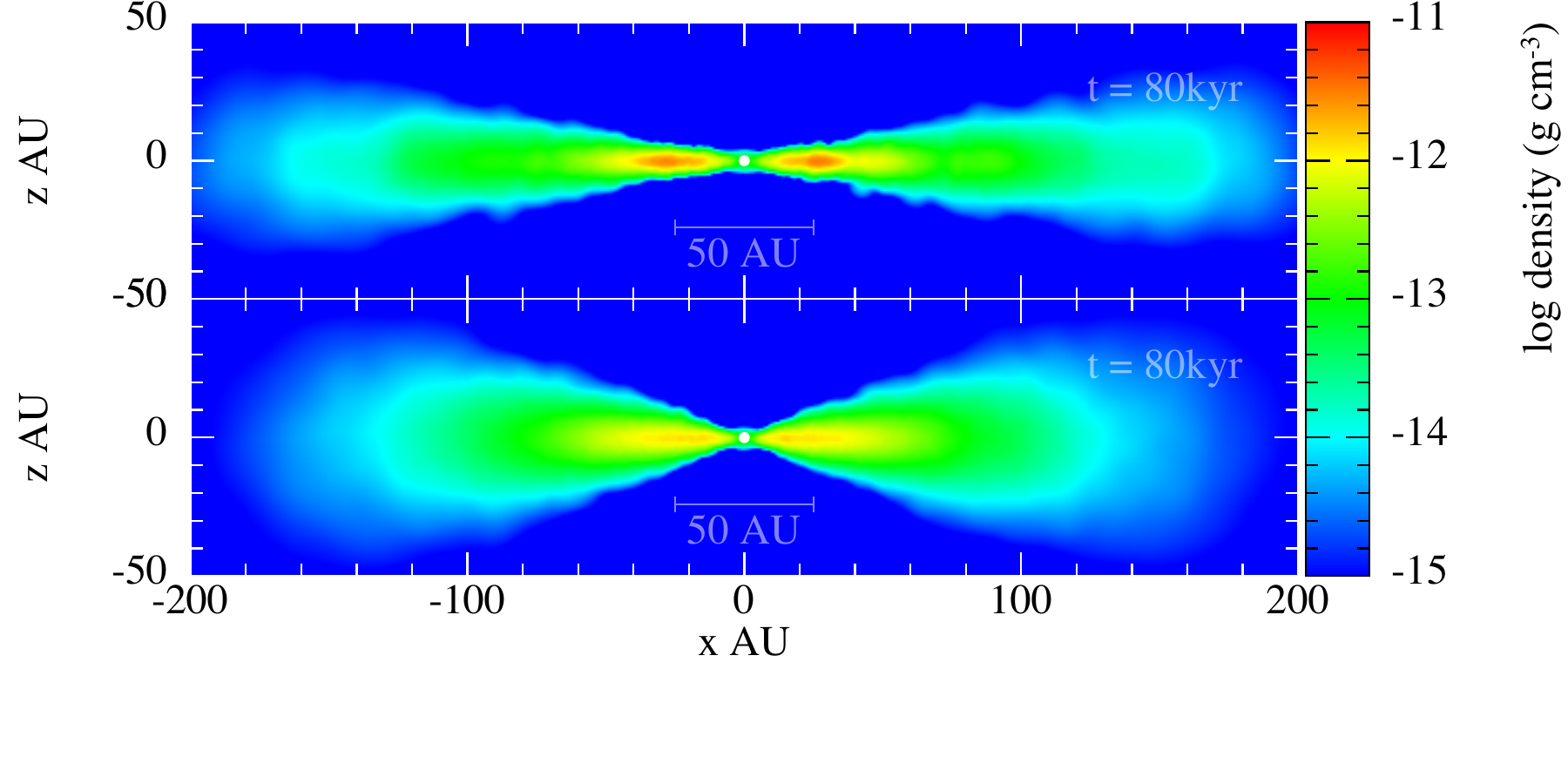}
    \caption{Cross-section of the density on the $x-z$ plane showing the disc  (edge-on view) at $t=15$ kyr, which forms around the protostar in run 9 (episodic accretion; top) and in run 11 (continuous accretion; bottom).}
    \label{fig:star-discs}
\end{figure}

In all simulations, the sink initially accretes vigorously with $\dot{M}_{\rm IAD} \sim 4 \times 10^{-5}$ M$_{\odot}$ yr$^{-1}$, with $\dot{M}_{\rm IAD}$ gradually decreasing over time. Each outburst is marked by a temporary drop in the sink’s accretion rate due to a decrease in the disc surface density (see Fig.~\ref{fig:star-discs}). During the quiescent accretion phases, the luminosity mostly remains within the range of $[0.1-1]$ L$_{\odot}$. However, during episodic accretion events, it spikes up. Higher values of $a_{\rm MRI}$ result in more intense episodic outbursts, a trend similar to that observed in forming BD (see Fig.~\ref{fig:a_mri_model}).

The continuous accretion model (run 11) differs from the episodic accretion models (runs 8–10) during low-mass star formation, in a manner similar to its behaviour during BD formation. Initially, the sink accretes at comparable rates across all simulations, but the accretion rate gradually declines as the available cloud core mass is depleted. Run 11 exhibits higher luminosity than models 8–10, except during outbursts, resulting in a hotter and thus also thicker disc, as can be seen in Fig.~\ref{fig:star-discs}. Nevertheless, the total energy emitted during each run is approximately the same, $E_{\rm total} \sim 10^{47}$ erg. Thus, we conclude that episodic accretion in low-mass star formation distributes the total emitted energy in a manner analogous to BD formation.

\begin{figure}[!htbp]
    \centering
    \includegraphics[width=\linewidth]{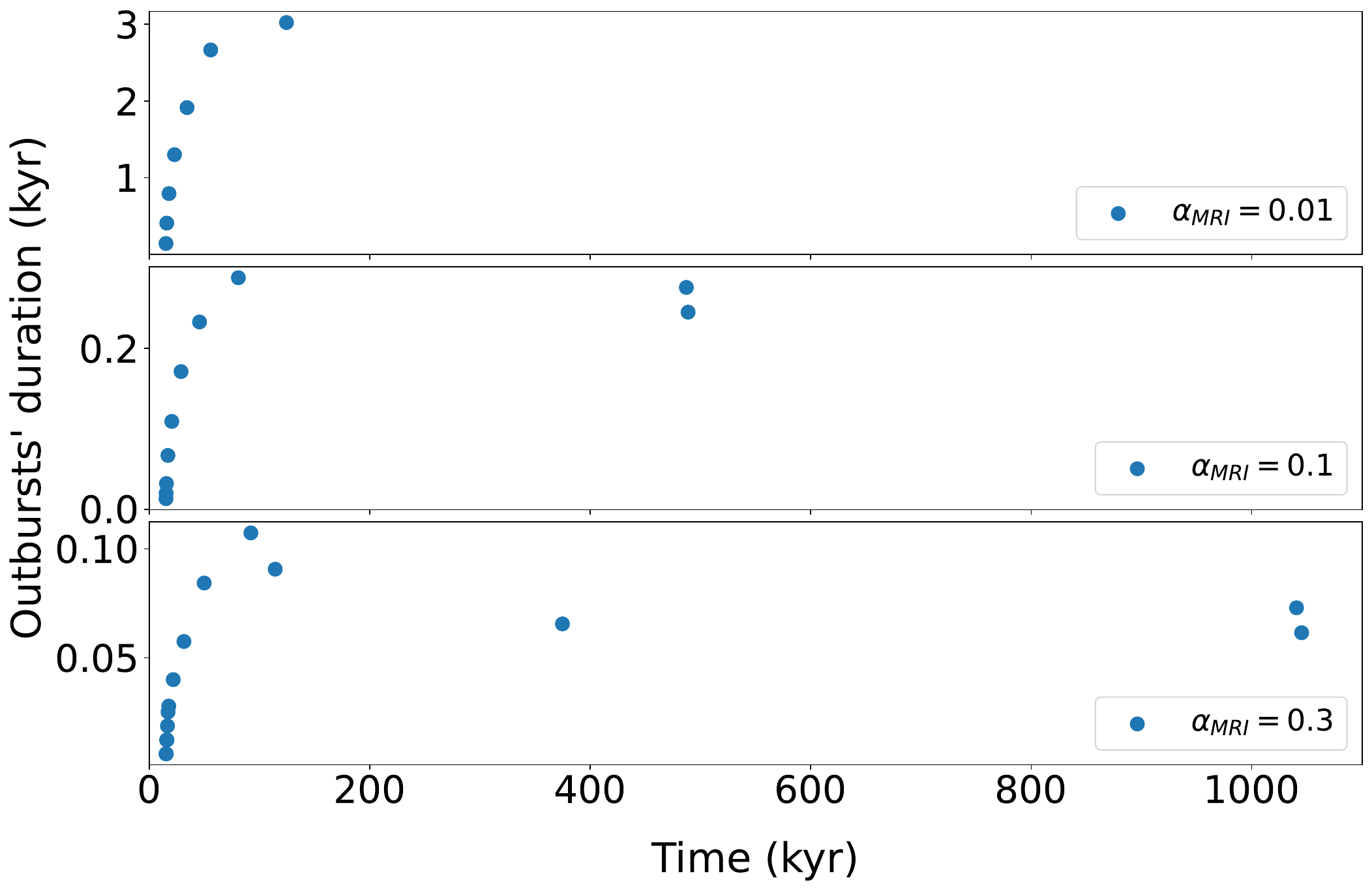}
    \caption{Effect of increasing $\alpha_{\rm MRI}$ on the duration of episodic outbursts for runs 8-10; see also Table~\ref{tab:runs_protostars}.}
    \label{fig:duration_same_bgr_protostar}
\end{figure}

\begin{figure}[!htbp]
    \centering
    \includegraphics[width=\linewidth]{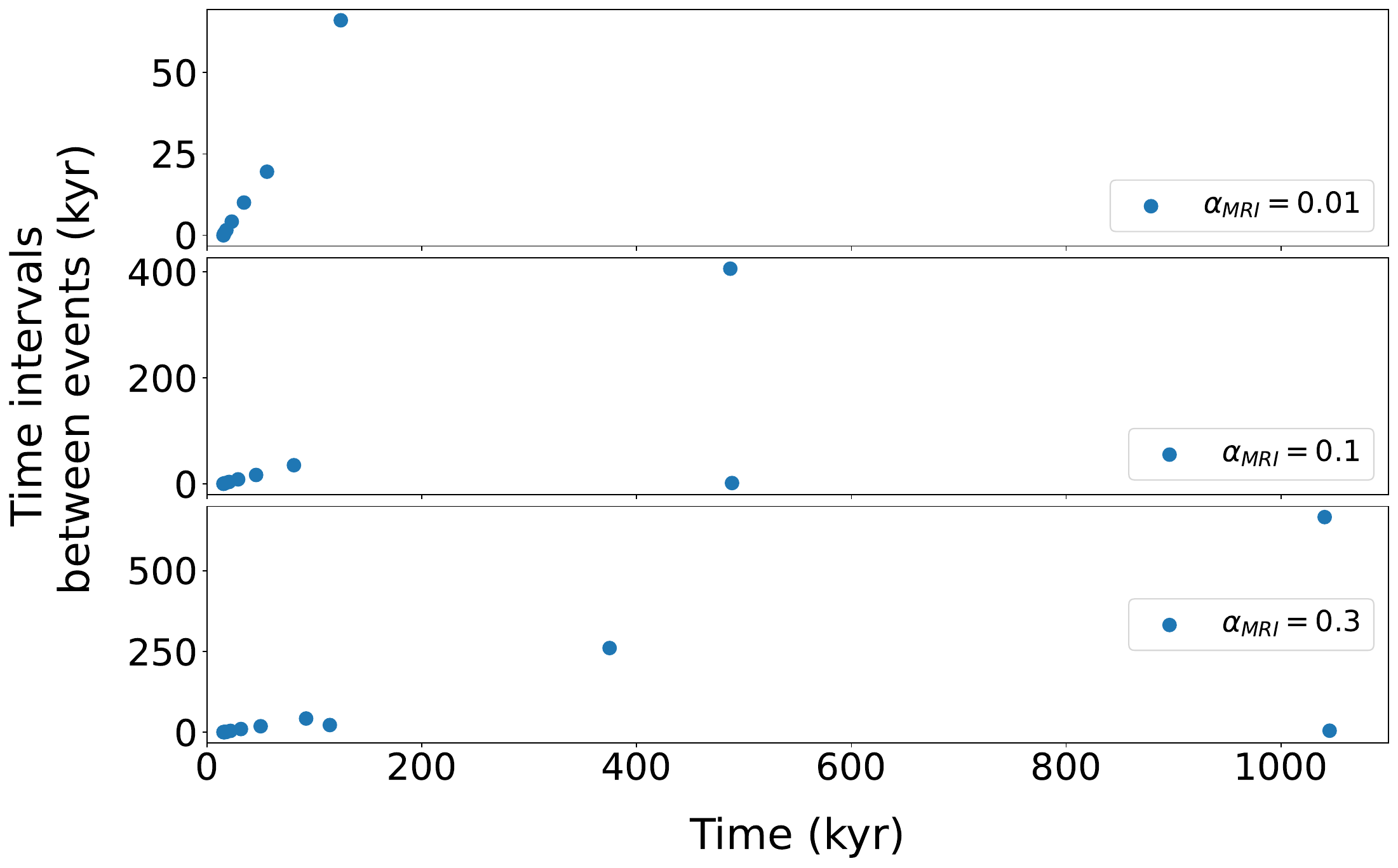}
    \caption{Effect of increasing $\alpha_{\rm MRI}$ on the time intervals between successive episodic outbursts for runs 8-10; see also Table~\ref{tab:runs_protostars}.}
    \label{fig:time_intervals_same_bgr_protostars}
\end{figure}

In Figures~\ref{fig:duration_same_bgr_protostar} and \ref{fig:time_intervals_same_bgr_protostars}, we present the time duration and time intervals between successive outbursts during the first $1.1$~Myr of the simulation, as no episodic event occurs afterwards. Higher values of $a_{\rm MRI}$ lead to shorter outbursts and shorter time intervals between successive events during low-mass star formation. For example, the shortest event lasts $\sim 144$ yr for $\alpha_{\rm  MRI} = 0.01$, whereas the longest outburst for $\alpha_{\rm MRI} = 0.3$ lasts only$\sim 107$ yr. Additionally, for $\alpha_{\rm  MRI} = 0.01$, the time interval between the third and the fourth event is $\sim 4217$ yr. However, for $\alpha_{\rm  MRI} = 0.1$ and $\alpha_{\rm  MRI} =0.3$, the time intervals between the third and the fourth events are $\sim 1316$ and $\sim 328$ yr, respectively. Thus, we observe a consistent trend for outbursts from low-mass stars and BDs, where higher $\alpha_{\rm MRI}$ values are associated with shorter event durations and intervals between successive events.

\subsection{Episodic accretion during the main accretion phase of low-mass star formation}

\begin{figure}[!htbp]
    \centering
    \includegraphics[width=\linewidth]{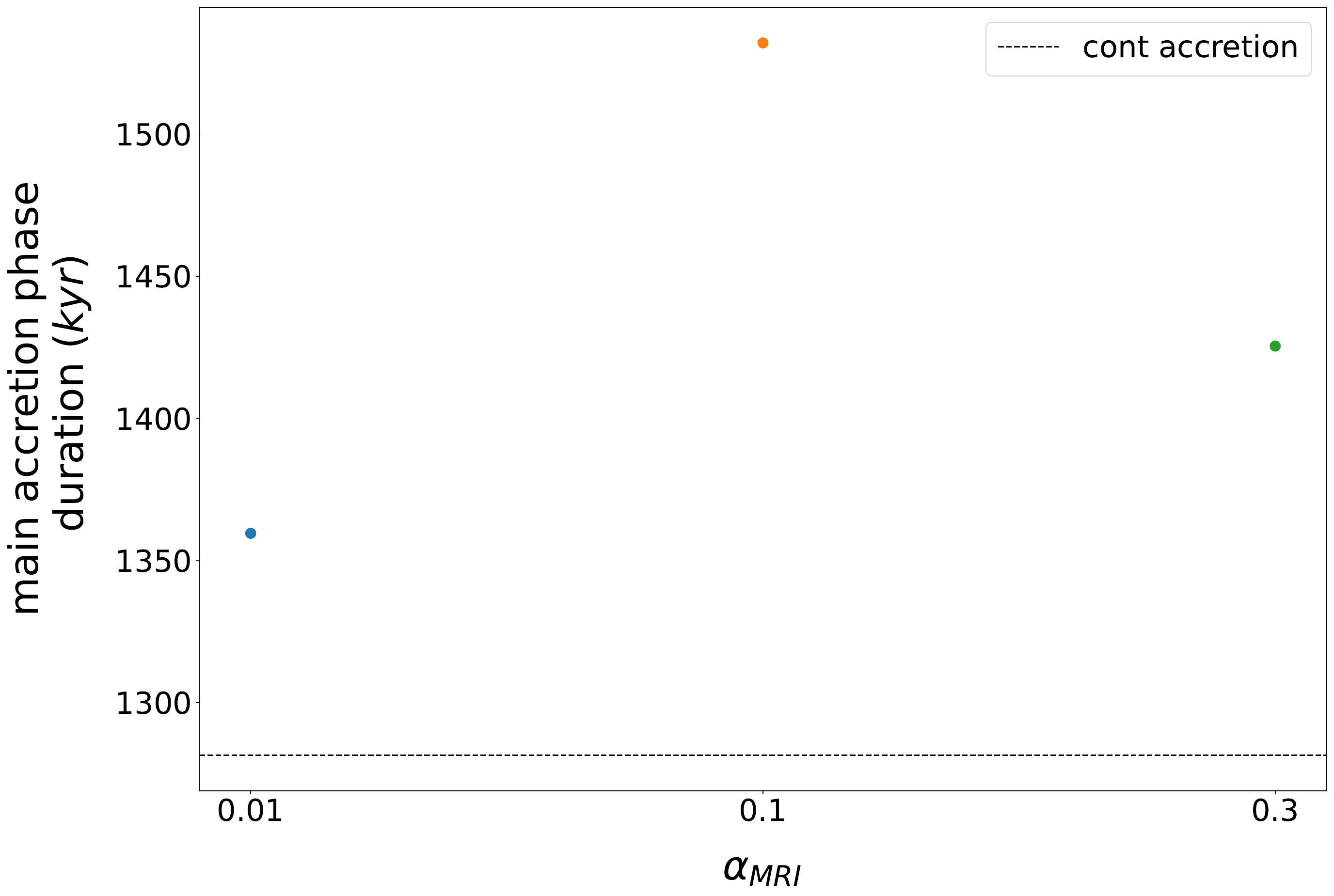}
    \caption{Duration of the main accretion phase for runs 8-11; see also Table~\ref{tab:runs_protostars}. The black horizontal dashed line corresponds to run 11 (continuous accretion).}
    \label{fig:pr_duration_main_acc}
\end{figure}

In Figure~\ref{fig:pr_duration_main_acc}, we illustrate the duration of the main accretion phase as a function of $\alpha_{\rm MRI}$ for the forming protostars. To ensure consistency in our comparison between the formation of proto-BDs and low-mass protostars, we apply the same criteria to define the main accretion phase as used in the proto-BD simulations. For $\alpha_{\rm MRI}=0.1$, the main accretion phase is the longest, while for $\alpha_{\rm MRI}=0.01$, the shortest. Additionally,  across all values of $\alpha_{\rm MRI}$, the episodic accretion models (runs 8–10) exhibit a longer main accretion phase than the continuous accretion model (run 11), which is the opposite of what seen in the BD formation simulations (see Fig.~\ref{fig:duration_main_acc_BDs}).

\begin{figure}[!htbp]
    \centering
    \includegraphics[width=\linewidth]{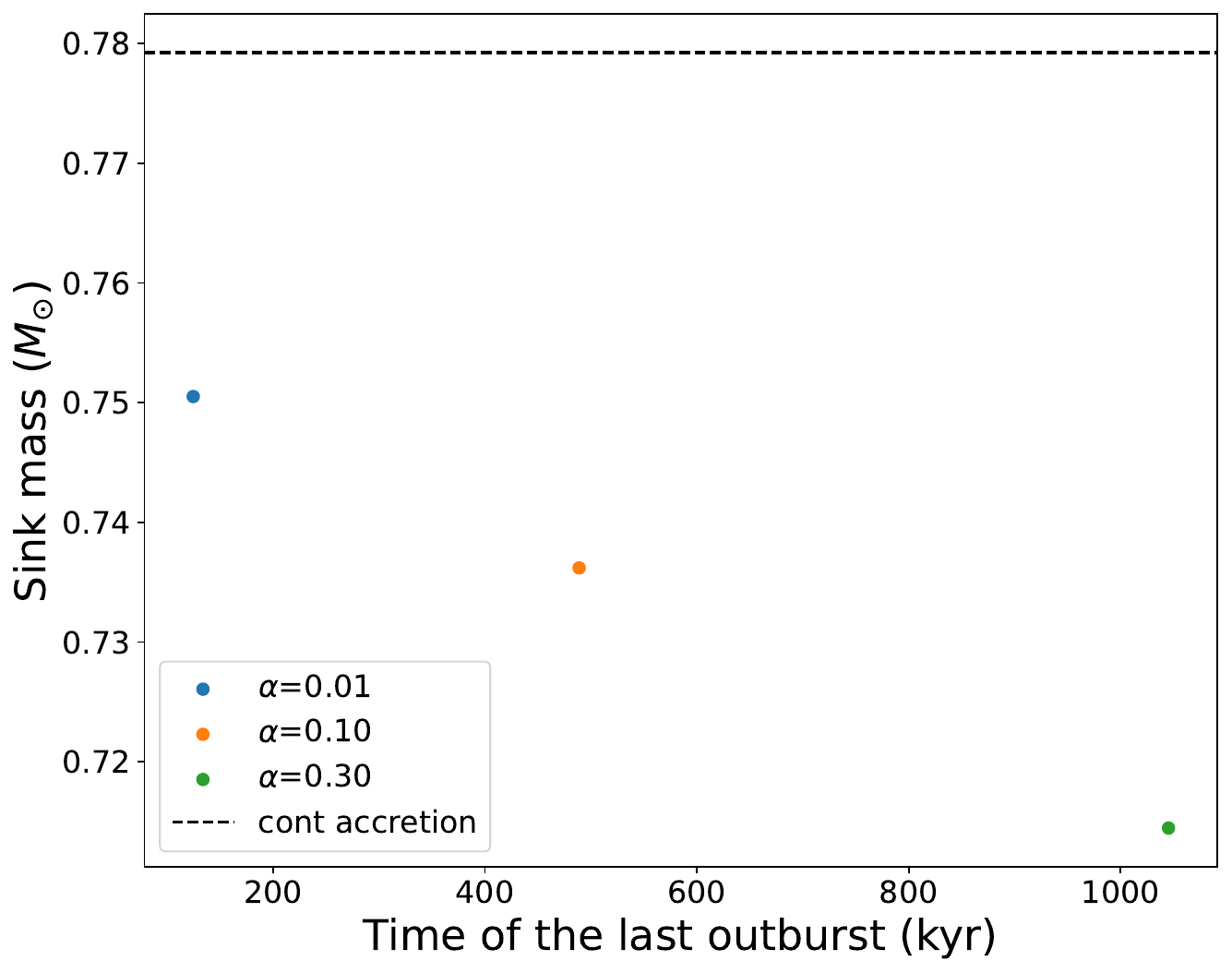}
    \caption{ Sink mass at the end of the main accretion phase vs. time of the last outburst for runs 8-11; see also Table~\ref{tab:runs_protostars}. The black horizontal dashed line corresponds to run 11 (continuous accretion).}
    \label{fig:pr_accreted_mass_main_acc}
\end{figure}

In Figure~\ref{fig:pr_accreted_mass_main_acc}, we show the sink mass at the end of the main accretion phases in relation to the time of the last episodic event, for runs 8-10. Notably, sink particles in the episodic accretion models accumulate more mass when the final episodic event occurs earlier. Moreover, in the continuous accretion simulation (run 11) the sink accretes more mass overall compared to the sink in the episodic accretion simulations (runs 8-10). In Table~\ref{tab:protostars_results} we summarize our results. We conclude that episodic accretion, accompanied by radiative outbursts, leads to lower gas accretion than continuous  accretion, similar to what is observed in BD formation.

A notable difference between proto-BDs and low-mass stars can be seen by comparing the sink accretion rates in the episodic and continuous cases (black vs orange lines) in 
Figs.~\ref{fig:a_mri_model} and \ref{fig:a_mri_protostars} (middle row). In contrast to BD formation, for low-mass star formation, following the last outburst, the accretion rate onto the sink swiftly converges into the one of the continuous accretion case. We see that the effect of episodic outbursts is more prominent in BD-mass cloud cores than in more massive cloud cores, presumably due to the lower gravitational potential energy of the BD-mass cores.  

\begin{table}[!htbp]
\caption{Summary of episodic accretion effects on low-mass star formation.}
  \centering
  \begin{tabularx}{\linewidth}{cccccc}
  \hline
  run & $t_{\rm main}$ & $t_{\rm final,event}$ & $M_{\rm sink,main }$ & $M_{\rm sink,1.6 Myr}$ & $\dot{M}_{\rm IAD,main}$ \\
  & (kyr) & (kyr) & (kyr) & (M$_{\odot}$) & (M$_{\odot}$ yr$^{-1}$)\\
  \hline \hline
  8  & 1359  &  124.6 & 0.7452 & 0.7589 & 4.8 $\times 10^{-8}$ \\
  9  &  1532 &  488.8 & 0.7282 & 0.7382 & 3.7 $\times 10^{-8}$\\
  10 &  1425 &  1045.2 & 0.7096 & 0.7195 & 4 $\times 10^{-8}$\\
  11 & 1226 & –  & 0.7740 & 0.7931 & 5.7 $\times 10^{-8}$\\
  \hline
  \end{tabularx}\label{tab:protostars_results}
\tablefoot{$t_{\rm main}$ is the duration of the main accretion phase, while $t_{\rm final,event}$ denotes the moment of the final outburst. $M_{\rm sink,main}$ is the sink mass at the end of the main accretion phase, and $M_{\rm sink,1.6 Myr}$ is the sink mass at the end of the simulation.  $\dot{M}_{\rm IAD,main}$ indicates the accretion rate into the sink at the end of the main accretion phase.
}
\end{table}

\section{Discussion}\label{sec:six}

The MRI viscosity $\alpha_{\rm MRI}$ is a free parameter in our models, but its value is uncertain. Therefore,  observations of outbursting proto-BDs and protostars are needed in order to constrain it. In Figures~\ref{fig:a_mri_model} and \ref{fig:a_mri_protostars}, the models for $\alpha_{\rm MRI} = 0.3$ result to maximum outburst luminosities reaching $\sim 10^3$ L${_\odot}$ and $\sim 10^4$ L${_\odot}$, respectively. \cite{2021A&A...647A..44V} presented the main characteristics of observed FUors and FUor-like objects \citep[see][Table A.1.]{2021A&A...647A..44V}; since there are no observed outbursts with $\geq 10^3$ L$_{\odot}$, values of $\alpha_{\rm MRI} \geq 0.3$ are probably unlikely. Values of $\alpha_{\rm MRI} \sim 0.1$ result predominantly to luminosities $\sim 10^1$ L$_{\odot}$, $\sim 10^2$ L$_{\odot}$ and outbursts duration on the order of $\sim 10^1$ yr, $\sim 10^2$ yr for proto-BDs and protostars, respectively, providing a better agreement with the observations. We note though that if outbursts with luminosities $\geq 10^3$ L$_{\odot}$ do occur, then their duration is $<20$ yr (see Figs.~\ref{fig:duration_same_bgr} and \ref{fig:duration_same_bgr_protostar}), and therefore statistically difficult to be observed.

\cite{2021MNRAS.501.3781R} presented a near-infrared study of accretion and outflow activity in six proto-BDs.  The bolometric luminosity, of their targets is in the range  $\sim 0.03 - 0.09$~L$_{\odot}$ and was used to derive mean accretion rates, which vary from $\sim 10^{-6}$ to $\sim 10^{-8}$ M$_{\odot}$ yr$^{-1}$. In the brown dwarf formation simulations, we notice that the last episodic event separates the evolution into high-accretion and low-accretion phases, notably for models 1 and 2, see Fig.~\ref{fig:a_mri_model}. This pattern is not apparent in low-mass star formation simulations, as the initial cloud core is more massive. During the high accretion phase, the accretion rate onto the proto-BD, $\dot{M}_{\rm IAD}$, ranges from $\sim 10^{-5}$ to $\sim 10^{-6}$ M$_{\odot}$ yr$^{-1}$, whereas during low-accretion from $\sim 10^{-7}$ to $\sim 10^{-9}$ M$_{\odot}$ yr$^{-1}$. The high-accretion period is short ($\leq 10-17$ kyr), making it statistically more difficult to observe. In contrast, it is plausible that we will frequently observe low-accretion rates, since low-accretion periods last longer. \cite{machida2009first} reach to a qualitatively similar conclusion. 

\subsection{ISO-OPH 200: a possible outbursting proto-BD?}

A detailed physical and chemical modelling of the ALMA observations for ISO-Oph 200 was conducted by \cite{riaz2021complex}. Their observational and model-derived signals are compatible with the predictions of the core collapse model for BD formation \citep{machida2009first} and they classify this object as a Class 0 proto-BD. 

They identify an infalling cloud core that merges onto a circumstellar pseudo-disk and an inner Keplerian pseudo-disk, and they  deduced two different outflows from the object; a wide-angled low-velocity CO molecular outflow ($\sim 1,000$~au), and a high-velocity (unresolved) [Fe II] atomic jet, with the former preceding the latter. The spatial scale of the CO outflow is comparable with the size of the cloud/pseudo-disk regions, resulting in the most extended CO molecular outflow identified to date.

\cite{riaz2021complex} from modelling  estimate a kinematic age of the system, i.e. the time since the system began to collapse to form the brown dwarf, of around $6 \pm 1$~kyr, and a dynamical age of the CO outflow, i.e.  the time since the outflow was launched by the proto-BD, of roughly $0.6 \pm 0.1$~kyr.

The presence of two different flows may indicate variable outflow activity. Additionally, the CO outflow's very young dynamical age, in comparison to the proto-BD system's kinematic age, implies a recent accretion event resulting in an outburst. This is corroborated  by the ALMA CO(2-1) spectrum, which shows a sudden rise in the flux by a factor of $\sim 20$ over just two velocity channels, and the observed spectral shape of a P-Cygni profile which is typically seen during accretion bursts  \citep{riaz2021complex}.

Our simulations support this argument.  More specifically, the kinematic age of the system is consistent with an episodic event occurring in the first few thousand years from the formation of the proto-BD. Furthermore, according to the simulations presented here, the dynamical age of the outflow ($\sim 0.6$~kyr) is comparable to what is expected for an outburst duration for $\alpha_{\rm MRI} \lesssim  0.01$. From Fig.~\ref{fig:duration_same_bgr} we see that the duration of the outbursts for $\alpha_{\rm MRI} \sim  0.01$ is on the order of a hundred years, while smaller $\alpha_{\rm MRI}$ will result in even longer outbursts. On the other hand, higher $\alpha_{\rm MRI}$ values lead to outbursts lasting only a few decades, thus they are not consistent with the dynamical age of the observed outflow.

\section{Conclusions}\label{sec:seven}

We capitalized the semi-analytic models of \cite{2009ApJ...694.1045Z,2010ApJ...713.1134Z}, which have been incorporated into hydrodynamic simulations of star formation by \cite{Stamatellos:2011a} to study episodic outbursts from proto-BDs. The model has two free parameters: the MRI viscosity parameter $\alpha_{\rm MRI}$, which regulates how quickly gas flows onto the proto-BD while the MRI is active, and $M_{\rm BGR}$, which reflects the proto-BD's quiescent accretion rate when the MRI is dormant. These parameters are weakly restricted and may differ among proto-BDs.

We performed  seven hydrodynamical simulations, all starting with an initial cloud mass of M$_{\rm cloud}=0.09~{\rm M}_{\odot}$, leading to the formation of objects within the brown dwarf mass range. These simulations included one continuous accretion model and six episodic accretion models, where we varied the free parameters, $\alpha_{\rm MRI}$ and $M_{\rm BGR}$. In addition, we ran four additional simulations using a more massive cloud, M$_{\rm cloud}=1.0~{\rm M}_{\odot}$, to investigate how episodic accretion affects brown dwarf formation and to examine the key differences and similarities compared to low-mass star formation. Our main conclusions are given below.

\begin{enumerate}
    \item The formation timescales for proto-BDs are at least one order of magnitude shorter than for protostars.
    \item During continuous and episodic accretion, the total amount of emitted energy is similar. However, episodic accretion redistributes the total emitted energy, allowing for long periods of low accretion luminosities and short luminosity outbursts. This behaviour is the same for forming proto-BDs and low-mass protostars.
    \item Almost all BD outbursts occur within $20$ kyr from the BD formation,  whereas the majority of outbursts from low-mass stars occur within $150$ kyr. Therefore, BD outbursts are more rare and hence more difficult to observe than outbursts from low-mass stars.
    \item In forming brown dwarfs, episodic accretion results to shorter main accretion phase than continuous accretion. However, the opposite is true in forming low-mass stars. Additionally, higher $\alpha_{\rm MRI}$ leads to a longer main accretion phase for brown dwarfs, but no such trend appears for low-mass stars. 
    \item Episodic accretion accelerates mass accretion onto to the forming object during the very early stages of its formation. However, both for brown dwarfs and low-mass stars, we find that episodic accretion results to less massive objects than continuous accretion. 
    \item We find a correlation between the object's mass at the end of the main accretion phase and the timing of the last episodic accretion event. For brown dwarfs, the later this final event occurs, the more massive the forming object is. However, for low-mass stars, we find the opposite trend: the later the final event, the less massive the forming object is.
    \item The effect of episodic outbursts is more prominent in BD-mass cloud cores than in more massive cloud cores. For brown dwarf forming cloud cores, the last outburst essentially splits the evolution into a short high-accretion and a much longer low-accretion phase. 
    
\end{enumerate}

\begin{acknowledgements}
We would like to thank the anonymous referee for their constructive review that has helped to improve the paper.
BR acknowledges funding from the Deutsche Forschungsgemeinschaft (DFG) - Projekt number RI-2919/2-3. DS acknowledges support from STFC grant ST/Y002741/1.
\end{acknowledgements}

% https://iopscience.iop.org/article/10.3847/1538-4357/aaba7b/pdf

% https://www.annualreviews.org/doi/pdf/10.1146/annurev.astro.34.1.207?casa_token=XpLckDfI7dsAAAAA:gM2TUpLRvQewTDFgvt608CbbpoK2pqwFafSgNCyebvuielNRDzEZrszo7MxFJmwf_-hd6BqDuEc

%https://www.researchgate.net/profile/Eduard-Vorobyov/publication/259743132_Episodic_Accretion_in_Young_Stars/links/02e7e52fe72ea2db66000000/Episodic-Accretion-in-Young-Stars.pdf

%\setcitestyle{numbers}
\bibliographystyle{aa}
\bibliography{references}

\begin{thebibliography}{63}
\expandafter\ifx\csname natexlab\endcsname\relax\def\natexlab#1{#1}\fi

\bibitem[{{Audard} {et~al.}(2014){Audard}, {{\'A}brah{\'a}m}, {Dunham},
  {Green}, {Grosso}, {Hamaguchi}, {Kastner}, {K{\'o}sp{\'a}l}, {Lodato},
  {Romanova}, {Skinner}, {Vorobyov}, \& {Zhu}}]{2014prpl.conf..387A}
{Audard}, M., {{\'A}brah{\'a}m}, P., {Dunham}, M.~M., {et~al.} 2014, in
  Protostars and Planets VI, ed. H.~{Beuther}, R.~S. {Klessen}, C.~P.
  {Dullemond}, \& T.~{Henning}, 387--410

\bibitem[{{Bae} {et~al.}(2014){Bae}, {Hartmann}, {Zhu}, \&
  {Nelson}}]{2014ApJ...795...61B}
{Bae}, J., {Hartmann}, L., {Zhu}, Z., \& {Nelson}, R.~P. 2014, \apj, 795, 61

\bibitem[{Balbus \& Hawley(1998)}]{RevModPhys.70.1}
Balbus, S.~A. \& Hawley, J.~F. 1998, Rev. Mod. Phys., 70, 1

\bibitem[{Balsara(1995)}]{balsara1995neumann}
Balsara, D.~S. 1995, J. Comput. Phys., 121, 357

\bibitem[{{Bate}(2011)}]{2011MNRAS.417.2036B}
{Bate}, M.~R. 2011, \mnras, 417, 2036

\bibitem[{Bate(2012)}]{bate2012stellar}
Bate, M.~R. 2012, \mnras, 419, 3115

\bibitem[{Bate {et~al.}(1995)Bate, Bonnell, \& Price}]{bate1995modelling}
Bate, M.~R., Bonnell, I.~A., \& Price, N.~M. 1995, \mnras, 277, 362

\bibitem[{Bell {et~al.}(1995)Bell, Lin, Hartmann, \& Kenyon}]{Bell:1995a}
Bell, K.~R., Lin, D. N.~C., Hartmann, L.~W., \& Kenyon, S.~J. 1995, \apj, 444,
  376

\bibitem[{Boley {et~al.}(2007)Boley, Hartquist, Durisen, \&
  Michael}]{boley2007internal}
Boley, A.~C., Hartquist, T.~W., Durisen, R.~H., \& Michael, S. 2007, \apjl,
  656, L89

\bibitem[{{Davidson} {et~al.}(2011){Davidson}, {Novak}, {Matthews}, {Matthews},
  {Goldsmith}, {Chapman}, {Volgenau}, {Vaillancourt}, \&
  {Attard}}]{2011ApJ...732...97D}
{Davidson}, J.~A., {Novak}, G., {Matthews}, T.~G., {et~al.} 2011, \apj, 732, 97

\bibitem[{{Dunham} {et~al.}(2015){Dunham}, {Allen}, {Evans},
  {Broekhoven-Fiene}, {Cieza}, {Di Francesco}, {Gutermuth}, {Harvey},
  {Hatchell}, {Heiderman}, {Huard}, {Johnstone}, {Kirk}, {Matthews}, {Miller},
  {Peterson}, \& {Young}}]{2015ApJS..220...11D}
{Dunham}, M.~M., {Allen}, L.~E., {Evans}, II, N.~J., {et~al.} 2015, \apjs, 220,
  11

\bibitem[{{Dunham} \& {Vorobyov}(2012)}]{2012ApJ...747...52D}
{Dunham}, M.~M. \& {Vorobyov}, E.~I. 2012, \apj, 747, 52

\bibitem[{{Evans} {et~al.}(2009){Evans}, {Dunham}, {J{\o}rgensen}, {Enoch},
  {Mer{\'\i}n}, {van Dishoeck}, {Alcal{\'a}}, {Myers}, {Stapelfeldt}, {Huard},
  {Allen}, {Harvey}, {van Kempen}, {Blake}, {Koerner}, {Mundy}, {Padgett}, \&
  {Sargent}}]{2009ApJS..181..321E}
{Evans}, II, N.~J., {Dunham}, M.~M., {J{\o}rgensen}, J.~K., {et~al.} 2009,
  \apjs, 181, 321

\bibitem[{{Fischer} {et~al.}(2023){Fischer}, {Hillenbrand}, {Herczeg},
  {Johnstone}, {Kospal}, \& {Dunham}}]{2023ASPC..534..355F}
{Fischer}, W.~J., {Hillenbrand}, L.~A., {Herczeg}, G.~J., {et~al.} 2023, in
  Astronomical Society of the Pacific Conference Series, Vol. 534, Protostars
  and Planets VII, ed. S.~{Inutsuka}, Y.~{Aikawa}, T.~{Muto}, K.~{Tomida}, \&
  M.~{Tamura}, 355

\bibitem[{Forgan {et~al.}(2009)Forgan, Rice, Stamatellos, \&
  Whitworth}]{forgan2009introducing}
Forgan, D., Rice, K., Stamatellos, D., \& Whitworth, A. 2009, \mnras, 394, 882

\bibitem[{Goodwin \& Whitworth(2007)}]{goodwin2007brown}
Goodwin, S.~P. \& Whitworth, A. 2007, \aap, 466, 943

\bibitem[{{Herczeg} {et~al.}(2017){Herczeg}, {Johnstone}, {Mairs}, {Hatchell},
  {Lee}, {Bower}, {Chen}, {Aikawa}, {Yoo}, {Kang}, {Kang}, {Chen}, {Williams},
  {Bae}, {Dunham}, {Vorobyov}, {Zhu}, {Rao}, {Kirk}, {Takahashi}, {Morata},
  {Lacaille}, {Lane}, {Pon}, {Scholz}, {Samal}, {Bell}, {Graves}, {Lee},
  {Parsons}, {He}, {Zhou}, {Kim}, {Chapman}, {Drabek-Maunder}, {Chung},
  {Eyres}, {Forbrich}, {Hillenbrand}, {Inutsuka}, {Kim}, {Kim}, {Kuan}, {Kwon},
  {Lai}, {Lalchand}, {Lee}, {Lee}, {Long}, {Lyo}, {Qian}, {Scicluna}, {Soam},
  {Stamatellos}, {Takakuwa}, {Tang}, {Wang}, \& {Wang}}]{Herczeg:2017a}
{Herczeg}, G.~J., {Johnstone}, D., {Mairs}, S., {et~al.} 2017, \apj, 849, 43

\bibitem[{Hubber {et~al.}(2011)Hubber, Batty, McLeod, \&
  Whitworth}]{hubber2011seren}
Hubber, D.~A., Batty, C.~P., McLeod, A., \& Whitworth, A.~P. 2011, \aap, 529,
  A27

\bibitem[{{Isella} {et~al.}(2009){Isella}, {Carpenter}, \&
  {Sargent}}]{2009ApJ...701..260I}
{Isella}, A., {Carpenter}, J.~M., \& {Sargent}, A.~I. 2009, \apj, 701, 260

\bibitem[{{Jankovic} {et~al.}(2021){Jankovic}, {Owen}, {Mohanty}, \&
  {Tan}}]{2021MNRAS.504..280J}
{Jankovic}, M.~R., {Owen}, J.~E., {Mohanty}, S., \& {Tan}, J.~C. 2021, \mnras,
  504, 280

\bibitem[{{Kadam} {et~al.}(2020){Kadam}, {Vorobyov}, {Reg{\'a}ly},
  {K{\'o}sp{\'a}l}, \& {{\'A}brah{\'a}m}}]{2020ApJ...895...41K}
{Kadam}, K., {Vorobyov}, E., {Reg{\'a}ly}, Z., {K{\'o}sp{\'a}l}, {\'A}., \&
  {{\'A}brah{\'a}m}, P. 2020, \apj, 895, 41

\bibitem[{{Kenyon} {et~al.}(1990){Kenyon}, {Hartmann}, {Strom}, \&
  {Strom}}]{1990AJ.....99..869K}
{Kenyon}, S.~J., {Hartmann}, L.~W., {Strom}, K.~M., \& {Strom}, S.~E. 1990,
  \aj, 99, 869

\bibitem[{{King} {et~al.}(2007){King}, {Pringle}, \&
  {Livio}}]{2007MNRAS.376.1740K}
{King}, A.~R., {Pringle}, J.~E., \& {Livio}, M. 2007, \mnras, 376, 1740

\bibitem[{{Lee} {et~al.}(2018){Lee}, {Kim}, {Myers}, {Saito}, {Kim}, {Kwon},
  {Lyo}, {Soam}, \& {Kim}}]{2018ApJ...865..131L}
{Lee}, C.~W., {Kim}, G., {Myers}, P.~C., {et~al.} 2018, \apj, 865, 131

\bibitem[{Lodato \& Rice(2004)}]{lodato2004testing}
Lodato, G. \& Rice, W. 2004, \mnras, 351, 630

\bibitem[{{Lomax} {et~al.}(2014){Lomax}, {Whitworth}, {Hubber}, {Stamatellos},
  \& {Walch}}]{Lomax:2014a}
{Lomax}, O., {Whitworth}, A.~P., {Hubber}, D.~A., {Stamatellos}, D., \&
  {Walch}, S. 2014, \mnras, 439, 3039

\bibitem[{{Lomax} {et~al.}(2015){Lomax}, {Whitworth}, {Hubber}, {Stamatellos},
  \& {Walch}}]{Lomax:2015a}
{Lomax}, O., {Whitworth}, A.~P., {Hubber}, D.~A., {Stamatellos}, D., \&
  {Walch}, S. 2015, \mnras, 447, 1550

\bibitem[{{MacFarlane} {et~al.}(2019{\natexlab{a}}){MacFarlane}, {Stamatellos},
  {Johnstone}, {Herczeg}, {Baek}, {Chen}, {Kang}, \& {Lee}}]{MacFarlane:2019a}
{MacFarlane}, B., {Stamatellos}, D., {Johnstone}, D., {et~al.}
  2019{\natexlab{a}}, \mnras, 487, 5106

\bibitem[{{MacFarlane} {et~al.}(2019{\natexlab{b}}){MacFarlane}, {Stamatellos},
  {Johnstone}, {Herczeg}, {Baek}, {Chen}, {Kang}, \& {Lee}}]{MacFarlane:2019p}
{MacFarlane}, B., {Stamatellos}, D., {Johnstone}, D., {et~al.}
  2019{\natexlab{b}}, \mnras, 487, 4465

\bibitem[{{MacFarlane} \& {Stamatellos}(2017)}]{MacFarlane:2017a}
{MacFarlane}, B.~A. \& {Stamatellos}, D. 2017, \mnras, 472, 3775

\bibitem[{Machida {et~al.}(2009)Machida, Inutsuka, \&
  Matsumoto}]{machida2009first}
Machida, M.~N., Inutsuka, S.-i., \& Matsumoto, T. 2009, \apjl, 699, L157

\bibitem[{{Machida} \& {Matsumoto}(2012)}]{2012MNRAS.421..588M}
{Machida}, M.~N. \& {Matsumoto}, T. 2012, \mnras, 421, 588

\bibitem[{{Mercer} \& {Stamatellos}(2017)}]{Mercer:2017a}
{Mercer}, A. \& {Stamatellos}, D. 2017, \mnras, 465, 2

\bibitem[{{Mercer} \& {Stamatellos}(2020)}]{Mercer:2020a}
{Mercer}, A. \& {Stamatellos}, D. 2020, \aap, 633, A116

\bibitem[{{Mercer} {et~al.}(2018){Mercer}, {Stamatellos}, \&
  {Dunhill}}]{Mercer:2018}
{Mercer}, A., {Stamatellos}, D., \& {Dunhill}, A. 2018, \mnras, 478, 3478

\bibitem[{{Morata} {et~al.}(2015){Morata}, {Palau}, {Gonz{\'a}lez}, {de
  Gregorio-Monsalvo}, {Ribas}, {Perger}, {Bouy}, {Barrado}, {Eiroa}, {Bayo},
  {Hu{\'e}lamo}, {Morales-Calder{\'o}n}, \&
  {Rodr{\'\i}guez}}]{2015ApJ...807...55M}
{Morata}, O., {Palau}, A., {Gonz{\'a}lez}, R.~F., {et~al.} 2015, \apj, 807, 55

\bibitem[{Morris \& Monaghan(1997)}]{morris1997switch}
Morris, J.~P. \& Monaghan, J.~J. 1997, J. Comput. Phys., 136, 41

\bibitem[{{Offner} {et~al.}(2010){Offner}, {Kratter}, {Matzner}, {Krumholz}, \&
  {Klein}}]{2010ApJ...725.1485O}
{Offner}, S. S.~R., {Kratter}, K.~M., {Matzner}, C.~D., {Krumholz}, M.~R., \&
  {Klein}, R.~I. 2010, \apj, 725, 1485

\bibitem[{Palau {et~al.}(2014)Palau, Zapata, Rodr{\'\i}guez, Bouy, Barrado,
  Morales-Calder{\'o}n, Myers, Chapman, Ju{\'a}rez, \& Li}]{palau2014ic}
Palau, A., Zapata, L.~A., Rodr{\'\i}guez, L.~F., {et~al.} 2014, \mnras, 444,
  833

\bibitem[{{Palla} \& {Stahler}(1993)}]{1993ApJ...418..414P}
{Palla}, F. \& {Stahler}, S.~W. 1993, \apj, 418, 414

\bibitem[{{Riaz} \& {Bally}(2021)}]{2021MNRAS.501.3781R}
{Riaz}, B. \& {Bally}, J. 2021, \mnras, 501, 3781

\bibitem[{Riaz {et~al.}(2017)Riaz, Brice{\~n}o, Whelan, \&
  Heathcote}]{riaz2017first}
Riaz, B., Brice{\~n}o, C., Whelan, E., \& Heathcote, S. 2017, \apjl, 844, 47

\bibitem[{Riaz \& Machida(2021)}]{riaz2021complex}
Riaz, B. \& Machida, M. 2021, \mnras, 504, 6049

\bibitem[{Riaz {et~al.}(2019)Riaz, Machida, \& Stamatellos}]{riaz2019alma}
Riaz, B., Machida, M., \& Stamatellos, D. 2019, \mnras, 486, 4114

\bibitem[{{Riaz} {et~al.}(2024){Riaz}, {Stamatellos}, \&
  {Machida}}]{2024MNRAS.529.3601R}
{Riaz}, B., {Stamatellos}, D., \& {Machida}, M.~N. 2024, \mnras, 529, 3601

\bibitem[{Riaz {et~al.}(2015)Riaz, Thompson, Whelan, \& Lodieu}]{riaz2015very}
Riaz, B., Thompson, M., Whelan, E., \& Lodieu, N. 2015, \mnras, 446, 2550

\bibitem[{{Shakura} \& {Sunyaev}(1973)}]{1973A&A....24..337S}
{Shakura}, N.~I. \& {Sunyaev}, R.~A. 1973, \aap, 24, 337

\bibitem[{{Shang} {et~al.}(2007){Shang}, {Li}, \&
  {Hirano}}]{2007prpl.conf..261S}
{Shang}, H., {Li}, Z.~Y., \& {Hirano}, N. 2007, in Protostars and Planets V,
  ed. B.~{Reipurth}, D.~{Jewitt}, \& K.~{Keil}, 261

\bibitem[{Shu {et~al.}(1987)Shu, Adams, \& Lizano}]{shu1987star}
Shu, F.~H., Adams, F.~C., \& Lizano, S. 1987, \araa, 25, 23

\bibitem[{Stamatellos \& Whitworth(2009)}]{stamatellos2009properties}
Stamatellos, D. \& Whitworth, A.~P. 2009, \mnras, 392, 413

\bibitem[{Stamatellos {et~al.}(2007)Stamatellos, Whitworth, Bisbas, \&
  Goodwin}]{stamatellos2007radiative}
Stamatellos, D., Whitworth, A.~P., Bisbas, T., \& Goodwin, S. 2007, \aap, 475,
  37

\bibitem[{{Stamatellos} {et~al.}(2011){Stamatellos}, {Whitworth}, \&
  {Hubber}}]{Stamatellos:2011a}
{Stamatellos}, D., {Whitworth}, A.~P., \& {Hubber}, D.~A. 2011, \apj, 730, 32

\bibitem[{Stamatellos {et~al.}(2012)Stamatellos, Whitworth, \&
  Hubber}]{stamatellos2012episodic}
Stamatellos, D., Whitworth, A.~P., \& Hubber, D.~A. 2012, \mnras, 427, 1182

\bibitem[{{Takakuwa} {et~al.}(2012){Takakuwa}, {Saito}, {Lim}, {Saigo},
  {Sridharan}, \& {Patel}}]{2012ApJ...754...52T}
{Takakuwa}, S., {Saito}, M., {Lim}, J., {et~al.} 2012, \apj, 754, 52

\bibitem[{{Vorobyov} \& {Basu}(2005)}]{2005ApJ...633L.137V}
{Vorobyov}, E.~I. \& {Basu}, S. 2005, \apjl, 633, L137

\bibitem[{{Vorobyov} \& {Basu}(2010)}]{2010ApJ...719.1896V}
{Vorobyov}, E.~I. \& {Basu}, S. 2010, \apj, 719, 1896

\bibitem[{{Vorobyov} \& {Basu}(2015)}]{2015ApJ...805..115V}
{Vorobyov}, E.~I. \& {Basu}, S. 2015, \apj, 805, 115

\bibitem[{{Vorobyov} {et~al.}(2021){Vorobyov}, {Elbakyan}, {Liu}, \&
  {Takami}}]{2021A&A...647A..44V}
{Vorobyov}, E.~I., {Elbakyan}, V.~G., {Liu}, H.~B., \& {Takami}, M. 2021, \aap,
  647, A44

\bibitem[{{Vorobyov} {et~al.}(2020){Vorobyov}, {Elbakyan}, {Takami}, \&
  {Liu}}]{Vorobyov:2020l}
{Vorobyov}, E.~I., {Elbakyan}, V.~G., {Takami}, M., \& {Liu}, H.~B. 2020, \aap,
  643, A13

\bibitem[{{Werner} {et~al.}(2004){Werner}, {Roellig}, {Low}, {Rieke}, {Rieke},
  {Hoffmann}, {Young}, {Houck}, {Brandl}, {Fazio}, {Hora}, {Gehrz}, {Helou},
  {Soifer}, {Stauffer}, {Keene}, {Eisenhardt}, {Gallagher}, {Gautier}, {Irace},
  {Lawrence}, {Simmons}, {Van Cleve}, {Jura}, {Wright}, \&
  {Cruikshank}}]{2004ApJS..154....1W}
{Werner}, M.~W., {Roellig}, T.~L., {Low}, F.~J., {et~al.} 2004, \apjs, 154, 1

\bibitem[{{Whitworth} \& {Stamatellos}(2006)}]{2006A&A...458..817W}
{Whitworth}, A.~P. \& {Stamatellos}, D. 2006, \aap, 458, 817

\bibitem[{{Zhu} {et~al.}(2009){Zhu}, {Hartmann}, \&
  {Gammie}}]{2009ApJ...694.1045Z}
{Zhu}, Z., {Hartmann}, L., \& {Gammie}, C. 2009, \apj, 694, 1045

\bibitem[{{Zhu} {et~al.}(2010){Zhu}, {Hartmann}, {Gammie}, {Book}, {Simon}, \&
  {Engelhard}}]{2010ApJ...713.1134Z}
{Zhu}, Z., {Hartmann}, L., {Gammie}, C.~F., {et~al.} 2010, \apj, 713, 1134

\end{thebibliography}
\end{document}